\documentclass[a4paper,fleqn,usenatbib]{mnras}

\usepackage[T1]{fontenc}
\usepackage{ae,aecompl}

\usepackage{graphicx}	% Including figure files
\usepackage{amsmath}	% Advanced maths commands
\usepackage{amssymb}	% Extra maths symbols
\usepackage{ulem}

\usepackage{newtxtext,newtxmath}

\usepackage{eso-pic}% http://ctan.org/pkg/eso-pic

\AddToShipoutPictureBG*{%
  \AtPageUpperLeft{%
    \hspace{0.75\paperwidth}%
    \raisebox{-3.5\baselineskip}{%
      \makebox[0pt][l]{\textnormal{DES-2020-0631}}
}}}%

\AddToShipoutPictureBG*{%
  \AtPageUpperLeft{%
    \hspace{0.75\paperwidth}%
    \raisebox{-4.5\baselineskip}{%
      \makebox[0pt][l]{\textnormal{FERMILAB-PUB-21-027-AE}}
}}}%

\title[OzDES Mg II Lags]{OzDES Reverberation Mapping Program: The first Mg II lags from five years of monitoring}

\author[Yu et al.]{
\parbox{\textwidth}{
\Large
Zhefu Yu,$^{1}$
Paul Martini,$^{1,2,3}$
A.~Penton,$^{4}$
T.~M.~Davis,$^{4,5}$
U.~Malik,$^{6}$
C.~Lidman,$^{5,7}$
B.~E.~Tucker,$^{5,6}$
R.~Sharp,$^{6}$
C.~S.~Kochanek,$^{1,2}$
B.~M.~Peterson,$^{1,2,8}$
M.~Aguena,$^{9,10}$
S.~Allam,$^{11}$
F.~Andrade-Oliveira,$^{10,12}$
J.~Annis,$^{11}$
J.~Asorey,$^{13}$
E.~Bertin,$^{14,15}$
D.~Brooks,$^{16}$
D.~L.~Burke,$^{17,18}$
J.~Calcino,$^{4}$
A.~Carnero~Rosell,$^{10,19,20}$
D.~Carollo,$^{5,21}$
M.~Carrasco~Kind,$^{22,23}$
M.~Costanzi,$^{24,25,26}$
L.~N.~da Costa,$^{10,27}$
M.~E.~S.~Pereira,$^{28}$
H.~T.~Diehl,$^{11}$
S.~Everett,$^{29}$
I.~Ferrero,$^{30}$
B.~Flaugher,$^{11}$
J.~Frieman,$^{11,31}$
J.~Garc\'ia-Bellido,$^{32}$
E.~Gaztanaga,$^{33,34}$
D.~W.~Gerdes,$^{28,35}$
D.~Gruen,$^{17,18,36}$
R.~A.~Gruendl,$^{22,23}$
J.~Gschwend,$^{10,27}$
G.~Gutierrez,$^{11}$
S.~R.~Hinton,$^{4}$
D.~L.~Hollowood,$^{29}$
D.~J.~James,$^{37}$
A.~G.~Kim,$^{38}$
R.~Kron,$^{11,31}$
K.~Kuehn,$^{7,39}$
N.~Kuropatkin,$^{11}$
G.~F.~Lewis,$^{40}$
M.~A.~G.~Maia,$^{10,27}$
M.~March,$^{41}$
J.~L.~Marshall,$^{42}$
F.~Menanteau,$^{22,23}$
R.~Miquel,$^{43,44}$
R.~Morgan,$^{45}$
A.~M\"oller,$^{46}$
A.~Palmese,$^{11,31}$
F.~Paz-Chinch\'{o}n,$^{22,47}$
A.~A.~Plazas,$^{48}$
E.~Sanchez,$^{13}$
V.~Scarpine,$^{11}$
S.~Serrano,$^{33,34}$
I.~Sevilla-Noarbe,$^{13}$
M.~Smith,$^{49}$
M.~Soares-Santos,$^{28}$
E.~Suchyta,$^{50}$
G.~Tarle,$^{28}$
D.~Thomas,$^{51}$
C.~To,$^{17,18,36}$
D.~L.~Tucker$^{11}$
\\
{\it \footnotesize (Affiliations listed at the end of the paper)}
}
}

\date{Accepted XXX. Received YYY; in original form ZZZ}

\pubyear{2021}

\begin{document}
\label{firstpage}
\pagerange{\pageref{firstpage}--\pageref{lastpage}}
\maketitle

\begin{abstract}
Reverberation mapping is a robust method to measure the masses of supermassive black holes (SMBHs) outside of the local Universe. Measurements of the radius -- luminosity ($R-L$) relation using the Mg II emission line are critical for determining these masses near the peak of quasar activity at $z \approx 1 - 2$, and for calibrating secondary mass estimators based on Mg II that can be applied to large samples with only single-epoch spectroscopy. We present the first nine Mg II lags from our five-year Australian Dark Energy Survey (OzDES) reverberation mapping program, which substantially improves the number and quality of Mg II lag measurements. As the Mg II feature is somewhat blended with iron emission, we model and subtract both the continuum and iron contamination from the multi-epoch spectra before analyzing the Mg II line. We also develop a new method of quantifying correlated spectroscopic calibration errors based on our numerous, contemporaneous observations of F-stars. The lag measurements for seven of our nine sources are consistent with both the H$\beta$ and Mg II $R-L$ relations reported by previous studies. Our simulations verify the lag reliability of our nine measurements, and we estimate that the median false positive rate of the lag measurements is $4\%$. 
\end{abstract}


%%%%%%%%%%%%%%%%%%%%%%%%%%%%%%%%%%%%%%%%%%%%%%%%%%

%%%%%%%%%%%%%%%%% BODY OF PAPER %%%%%%%%%%%%%%%%%%

%Introduction 
\section{Introduction} \label{sec:intro}
Massive galaxies ubiquitously host supermassive black holes (SMBHs) at their centers \citep[e.g.,][]{Richstone1998,Kormendy2013}, and SMBH mass measurements are fundamental to the study of the evolution of active galactic nuclei (AGNs) and their host galaxies. Studies have measured SMBH masses in the local Universe using stellar and gas kinematics \citep[e.g.,][]{Kormendy1995,Gebhardt2009,Barth2016}. However, these methods require observations of stars or gas that resolve the black hole's region of influence, so they can only be applied to nearby galaxies with little accretion. The GRAVITY project applied the dynamical method to the quasar 3C 273 using the Very Large Telescope Interferometer \citep[e.g.,][]{Gravity2018} and showed the potential of extending this method to higher redshifts in the future. 

Outside of the local Universe, a robust method to measure SMBH masses is the reverberation mapping (RM) technique \citep{Blandford1982,Peterson1993}. RM maps the response of the broad emission line variability to the continuum variability of AGNs. The mean time lag $\langle \tau \rangle$ between the continuum and line lightcurves corresponds to the typical light travel time from the accretion disk to the broad line region (BLR). The black hole mass is determined from the virial theorem
\begin{equation}
M_{\rm BH} = \frac{f R \Delta v^2}{G}
\label{eq:bhmass}
\end{equation}
where $f$ is a dimensionless ``virial factor'' that accounts for the BLR structure and kinematics, $R = c \langle \tau \rangle$ is the typical BLR size, and $\Delta v$ is the broad line width. RM studies with high cadence and signal-to-noise ratio (SNR) data \citep[e.g.,][]{Grier2013Peterson,Horne2020} can further resolve the ``transfer function'' $\Psi(v,\tau)$ which characterizes the line variability as a function of the line-of-sight velocity $v$ and the time delay $\tau$. In the linearized response model, the varying component of the emission line is the convolution of the varying component of the continuum with the transfer function $\Psi(v,\tau)$.

RM studies have found a correlation between the continuum luminosity $L$ and the BLR size $R$ based on the lags of several broad lines, especially H$\beta$ \citep[e.g.,][]{Kaspi2000,Bentz2009,Bentz2013,Grier2017}, but also Mg II \citep[e.g.,][]{Shen2016,Czerny2019,Homayouni2020,Zajacek2020} and C IV \citep[e.g.,][]{Kaspi2007,Lira2018,Hoormann2019,Grier2019}, while some studies found deviations from the $R - L$ relations that correlated with, e.g., the accretion rate and the spectral energy distribution \citep[e.g.,][]{Du2016,Fonseca2020,Dalla2020}. The $R - L$ relation is important because it enables black hole mass estimates for large samples using only single-epoch spectra. This is especially valuable for studies of the SMBH population in AGNs \citep[e.g.,][]{Vestergaard2008,Kelly2013}. 

The Mg II $R - L$ relation is particularly relevant to single-epoch mass estimates at redshift $\sim 1 - 2$ where the Mg II line can be observed in optical spectra, and where AGN activity peaks \citep[e.g.,][]{Wolf2003,Ueda2014}. Until recently, there were less than a dozen Mg II lag measurements \citep{Metzroth2006,Shen2016,Lira2018,Czerny2019}. \citet{Homayouni2020} presented a larger Mg II lag sample from the Sloan Digital Sky Survey (SDSS) RM project \citep[e.g.,][]{Shen2015}, which showed a larger scatter in the Mg II $R - L$ relation than for H$\beta$. One explanation for this scatter was proposed by \citet{Guo2020}, who predicted that Mg II responds weakly to changes in the continuum luminosity based on photoionization models. However, their conclusions for the $R - L$ relation were qualitative and it is unclear whether this effect could account for all of the observed scatter. Additional measurements are needed to better determine the Mg II $R - L$ relation and its intrinsic scatter. 

It is now possible to increase the RM sample size relatively quickly by using multi-fiber spectroscopic surveys instead of single-slit spectrographic campaigns. However, while fiber instruments enable simultaneous monitoring of many AGN, the flux calibration for fiber spectra is more challenging than for slit spectra. This is because more factors can affect the flux in the aperture, especially wavelength-dependent fiber losses and variations in instrument throughput and alignment between the fibers that observe the AGN and calibration sources \citep[e.g.,][]{Hoormann2019}. The methods of calibrating fiber spectra vary among different surveys. For example, the SDSS RM project uses the $r$-band photometry of F-stars observed along with the AGNs to calibrate the spectra. They then use optimized model fits to improve the calibration based on the assumption that the fluxes of the narrow emission lines are constant with time \citep[e.g.,][]{Shen2015,Shen2016}. The Australian Dark Energy Survey (OzDES) project calibrates the spectra using multi-band photometry of each quasar from the Dark Energy Survey (DES) obtained at nearly the same epoch as each spectroscopic observation \citep{King2015,Hoormann2019}. These calibration approaches can introduce correlated errors in the wavelength-dependent flux calibration, which complicates the continuum modeling and the line flux measurements. 

Multi-object RM campaigns generally have fewer epochs and lower SNR than campaigns that intensively monitor individual sources. For example, the AGN Space Telescope and Optical RM project monitored NGC 5548 using the Hubble Space Telescope \citep[e.g.,][]{DeRosa2015}. They obtained 171 approximately daily spectra with SNR $\sim100$. In contrast, the spectroscopic cadence of the SDSS RM project was about four days during the first year, every two weeks from 2015 to 2017 and every month during the following years with spectral SNR $\sim15$ for the Mg II line \citep[e.g.,][]{Homayouni2020}. The OzDES RM program obtains about monthly spectroscopy with the SNR $\sim20$ \citep[e.g.,][]{King2015,Hoormann2019}. The low cadence and SNR of the multi-object RM campaigns lead to greater challenges in assessing the lag detection and the uncertainties. 

In this paper we present the first Mg II lags from the OzDES RM program based on five years of monitoring data. We provide a brief overview of the DES and the OzDES observations in Section \ref{sec:observations}. We then introduce the modeling of the continuum and the iron emission in Section \ref{subsec:ironfit} and discuss the correlated errors from the spectroscopic calibration in Section \ref{subsec:calerr}. Section \ref{sec:lcanl} presents the lag measurements and the quality cuts. We describe the simulations for assessing the lag validity in Section \ref{sec:simulations}. Finally, in Section \ref{sec:RL_BHMass} we present our black hole mass measurements and discuss the  $R - L$ relation. Section \ref{sec:summary} is a brief summary of the paper. Throughout this work we adopt a $\Lambda$CDM cosmology with $H_0 = 70 \, {\rm km/s/Mpc}$, $\Omega_m = 0.3$, $\Omega_{\Lambda} = 0.7$.

%Observations 
\section{Observations} \label{sec:observations}
DES is a ground-based, wide-area imaging survey \citep{DESDR1}. The survey used the $2.2^{\circ}$ diameter field of view Dark Energy Camera \citep[DECam, ][]{DECam} with the $grizY$-band filters on the 4-m Victor M. Blanco telescope at the Cerro Tololo Inter-American Observatory. DES began a 6-yr survey in 2013 that both mapped a deep, wide $5000 \, {\rm deg}^2$ region and monitored 10 supernova (SN) fields ($27 \, {\rm deg}^2$ total) with an approximately weekly cadence each year between mid-August and mid-February. 

OzDES is a spectroscopic survey that follows up the targets in the DES SN fields \citep[e.g.,][]{Yuan2015,Childress2017,Lidman2020}. OzDES uses the AAOmega spectrograph \citep{AAOmega} with the Two Degree Field (2dF) multi-fiber positioner \citep{2dF} on the 4-m Anglo-Australian Telescope (AAT). The OzDES RM program monitors 771 quasars at $0.1 < z < 4.5$ in the DES SN fields with approximately monthly spectroscopy and combines these data with the approximately weekly photometry from DES. OzDES observed these fields through most of the DES program, and has continued some longer-term monitoring. In previous work, we have used the photometric data to characterize the sizes of the accretion disks \citep{Mudd17,Yu2020_PhotRM}. In this paper we analyze the first five years of photometric and spectroscopic data.

We start our analysis with 470 quasars from the OzDES RM sample at $0.65 < z < 1.92$ where the spectra fully cover the Mg II line. We calibrate the spectra using the pipeline presented by \citet{Hoormann2019}. Briefly, the pipeline fits a second-order polynomial to the scaling factors from the instrumental $gri$ magnitudes derived from convolution of the DES filters with the spectroscopic data to the $gri$ photometric data from the nearest DES epoch. We calculate this calibration polynomial for each spectroscopic epoch. Since previous observations indicate that the Mg II lags are similar to the H$\beta$ lags \citep[e.g.,][]{Clavel1991,Metzroth2006,Homayouni2020}, we attempted to select sources whose expected observed-frame H$\beta$ lags $\tau_{{\rm H\beta,exp}}$ satisfy $\tau_{{\rm H\beta,exp}} < 0.35 {\rm yr}$, $0.75 {\rm yr} < \tau_{{\rm H\beta,exp}} < 1.25 {\rm yr}$ or $1.75 {\rm yr} < \tau_{{\rm H\beta,exp}} < 2.25 {\rm yr}$, to avoid the seasonal gaps. This reduced our sample to 203 sources. The selection was based on a table of expected lags from the start of the project at 2012 which used the \citet{Bentz2009} $R-L$ relation instead of the more recent \citet{Bentz2013} relation we use elsewhere in the paper. In addition, we discovered that the expected lag table had an error, which is why our sample includes some quasars with lags in the seasonal gaps.

%Spectroscopic Analysis
\section{Spectroscopic Analysis} \label{sec:specanl}

%Continuum & Iron subtraction
\subsection{Continuum \& Iron Subtraction} \label{subsec:ironfit}

In addition to the standard problem of estimating the continuum, a key challenge with the measurement of Mg II in quasars is the presence of many strong iron emission lines in the immediate vicinity of the Mg II line, including some that are blended \citep[e.g.,][]{Wills1980,Wills1985,Verner1999}. The broad iron lines may also reverberate and contaminate the Mg II lag signal \citep[e.g.,][]{Barth2013}. We consequently developed a Markov chain Monte-Carlo (MCMC) tool that simultaneously identifies the best model combination of a broadened iron template and a power law continuum, as well as estimates the uncertainties. The model is described by  
\begin{subequations}
\begin{align}
& f_{\rm model}(\lambda) = f_{\rm c}(\lambda) + f_{\rm Fe}(\lambda) \\
& f_{\rm c}(\lambda) = A_{\rm c} (\lambda/\lambda_0)^{\alpha} \\
& f_{\rm Fe}(\lambda) = A_{\rm t} f_{\rm t}(\lambda) \circledast G(w) 
\end{align}
\label{eq:ironmodel}%
\end{subequations}
where $f_{\rm c}$ and $f_{\rm Fe}$ are the continuum flux and the iron flux, $A_{\rm c}$ and $\alpha$ are the amplitude and slope of the power-law, $\lambda_0 = 2599$ \AA\, is a constant wavelength, $f_{\rm t}$ is the iron template, $A_{\rm t}$ is the template amplitude, and $G(w)$ is a Gaussian kernel with a width of $w$. The Gaussian kernel accounts for the velocity broadening of the iron emission region. The model has four free parameters: $A_{\rm c}$, $\alpha$, $A_{\rm t}$ and $w$. 

\begin{figure}
\includegraphics[width=\linewidth]{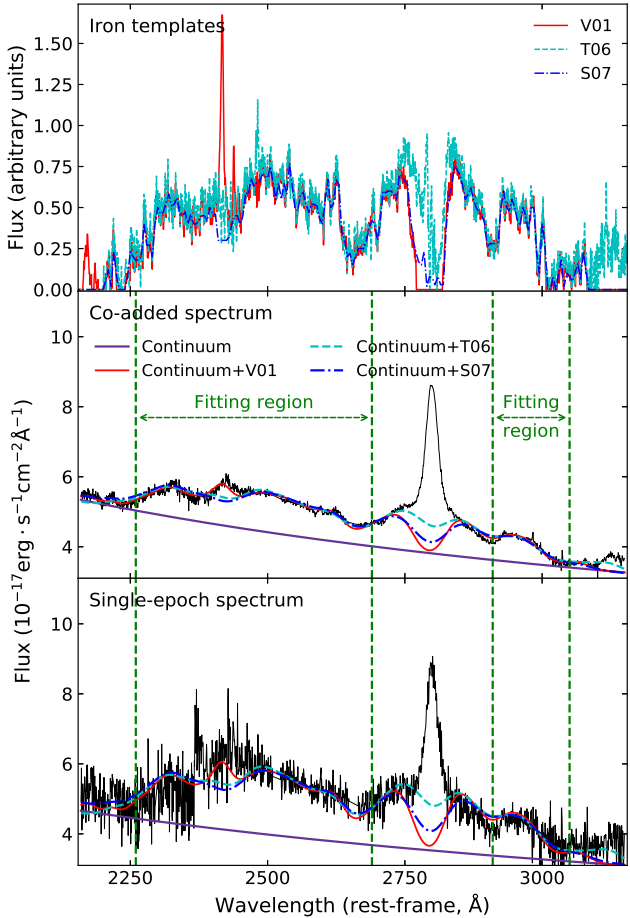}
\caption{Illustration of our fitting process. ({\it Top panel}) The V01 (red), T06 (cyan) and S07 (blue) iron templates. ({\it Middle and bottom panels}) Examples of the continuum and iron fit for DES J003052.76$-$430301.08. The middle and bottom panels show the co-added spectra and one of the single-epoch spectra, respectively. The green dashed lines mark the continuum $+$ iron fitting regions. The purple solid line represents the best-fit continuum. The red solid, cyan dashed and blue dash-dotted lines represent the best fit continuum $+$ iron models using the V01, T06 and S07 iron templates, respectively. All spectra have been shifted to the rest frame.}
\label{fig:ironfit}
\end{figure}

We adopt the iron templates from \citet[][V01 hereafter]{Vestergaard2001}, \citet[][T06 hereafter]{Tsuzuki2006} and \citet[][S07 hereafter]{Salviander2007}. These are empirical templates derived from the Seyfert galaxy I Zwicky 1. V01 artificially set the template flux to zero around the Mg II line where it is hard to separate the iron emission from the Mg II emission. T06 uses synthetic spectra from photoionization models to separate the iron emission from the Mg II line and uses the model to fill in the Mg II line region. S07 estimates the iron emission around the Mg II line using the theoretical iron model from \citet{Sigut2003}. Figure \ref{fig:ironfit} compares these three iron templates in the top panel. The templates are in general agreement with one another, except for differences near the Mg II line. In addition, only the V01 template includes the Fe III feature at $\sim$2400 \AA.\,

There is a potentially significant degeneracy between the continuum and iron emission amplitude. Fitting the spectra over a wider wavelength range can reduce the degeneracy between the continuum and the iron emission. However, a wider region may not be spanned by all the iron templates and increases the calculation expense. As a trade-off, we fit the quasar spectra over two rest-frame wavelength ranges: 2260 - 2690 \AA\, and 2910 - 3050 \AA.\, We do not fit the spectra near the Mg II line since the strong line emission would increase the difficulty of the continuum-iron separation. In addition, the broad line shape is not expected to be well fitted by the same simple, parametric function at every epoch. 

We calculate the Mg II line flux over the rest-frame wavelength range of 2700 - 2900 \AA.\, For each step of the MCMC chain, we integrate $f_{\rm model}(\lambda)$ over this line region to calculate the continuum + iron flux $F_{\rm model}$. We adopt the median of the MCMC chain as the best-fit model flux $\langle F_{\rm model} \rangle$ and adopt half the difference between the 16th and 84th percentile as the model flux uncertainty $\sigma_{\rm model}$. We then calculate the line flux $F_{\rm line}$ and uncertainty $\sigma_{\rm line}$ as
\begin{subequations}
\begin{align}
& F_{\rm line} = F_{\rm total} - \langle F_{\rm model} \rangle\\
& \sigma_{\rm line}^2 = \sigma_{\rm total}^2 + \sigma_{\rm model}^2
\end{align}
\label{eq:lineflux}%
\end{subequations}
where $F_{\rm total}$ is the integrated flux over the line region without continuum subtraction and $\sigma_{\rm total}$ is the uncertainty of $F_{\rm total}$. We calculate $F_{\rm total}$ and $\sigma_{\rm total}$ as 
\begin{subequations}
\begin{align}
& F_{\rm total} = \sum_i f_{{\rm total},i} \Delta \lambda \quad {\rm and} \\
& \sigma_{\rm total} = \sqrt{\sum_i \sigma_{{\rm total},i}^2} \Delta \lambda 
\end{align}
\label{eq:ftoterr}%
\end{subequations}
where $f_{{\rm total},i}$ and $\sigma_{{\rm total},i}$ are the flux and uncertainty of the $i^{\rm th}$ pixel and $\Delta \lambda$ is the pixel size. 

We are unable to well constrain the velocity broadening width $w$ for the single-epoch spectra with low SNR. We therefore use the co-added spectra to constrain the width $w$ for each source and then fit the single-epoch spectra with $w$ fixed. This assumes that the velocity broadening of the iron features does not change between epochs. In principle the broad emission lines can have a ``breathing mode'' where the line width changes as the continuum varies \citep[e.g.,][]{Korista2004,Guo2020,Wang2020}. Assuming $R \propto L^{0.5}$ according to previous H$\beta$ RM studies \citep[e.g.,][]{Bentz2009,Bentz2013}, the virial theorem predicts $w \propto L^{-0.25}$. The optical continuum variability is on the order of a few percent \citep[e.g.,][]{Peterson2001}. Assuming a $10\%$ continuum variability, the broadening width $w$ would vary by $2.5\%$. If we vary the broadening width $w$ by $2.5\%$ when calculating the Mg II line flux, the changes in the line flux are generally much smaller than the line flux uncertainties. While broad lines can have different breathing behaviours, the breathing is generally weaker than the $w \propto L^{-0.25}$ correlation \citep[e.g.,][]{Guo2020,Wang2020}, so the discussion above provides a conservative estimate. Therefore, the variability of the broadening width does not have a significant impact on our Mg II line flux measurements. 

In some sources there are narrow absorption doublets within the fitting region that could be due to interstellar or intergalactic absorption, or due to the instrumental artifacts. We linearly interpolate over these regions and increase the uncertainties of the affected pixels by a factor of 10. This decreases the inverse variance of these pixels by a factor of 100 so that they have negligible impact on the fits. Figure \ref{fig:ironfit} shows an example of the process. The models are a good match to the observed spectra and the three iron templates give similar fits to the spectra except in the Mg II line region. The V01 and S07 models do not match the spectra well near the right boundary since they have no coverage beyond 3090 \AA.\, As both the continuum and the iron emission vary over time, the model amplitudes differ between the co-added spectra and the single-epoch spectra. We show corner plots for the model parameters in the appendix. Additional figures for the other epochs and other sources in the main sample and the root-mean-square (rms) spectra are available in the online journal.

%Calibration uncertainties
\subsection{Calibration Uncertainty} \label{subsec:calerr}

\begin{figure*}
\includegraphics[width=\linewidth]{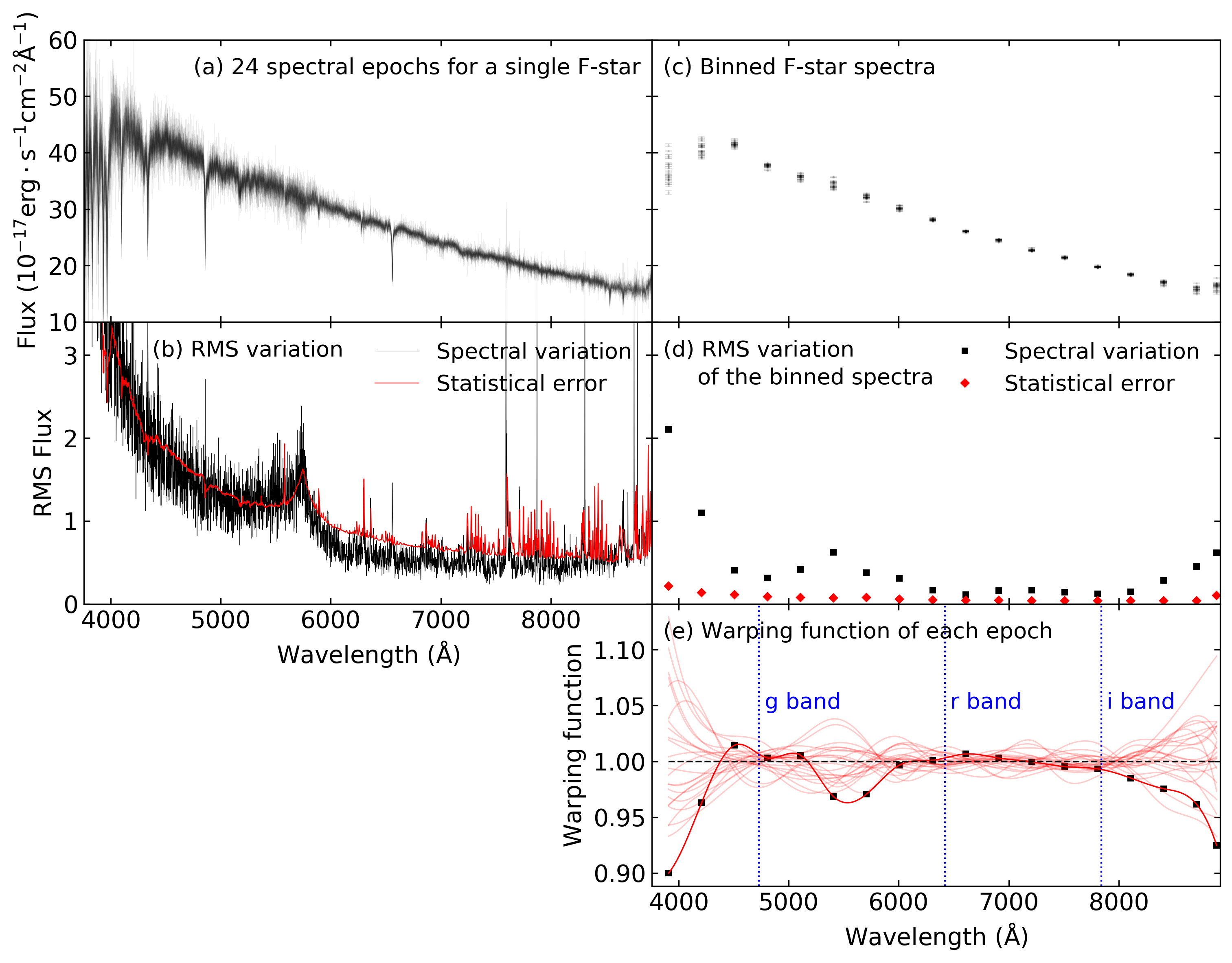}
\caption{Illustration of the process for estimating the calibration uncertainties. (a) Spectra of the F-star DES J022526.89$-$044520.09 at 24 epochs. (b) rms variation in the flux computed from the 24 epochs (black lines) and the mean statistical error (red line). The units are the same as panel (a). (c) Binned spectra of panel (a) with a bin size of 300 \AA.\, (d) Binned version of panel (b). The red dots show the dramatic reduction in statistical errors after binning. The black dots reveal the variation that remains in the binned data that are inconsistent with Poisson errors alone. (e) Warping functions. Each warping function corresponds to an epoch of the F-star spectra. The black squares represent the discrete warping function of the first epoch. The bold red line represents the continuous warping function of this epoch from the spline interpolation. The lighter red lines show the warping functions for the other epochs. The black dashed line is drawn at unity. The blue dotted lines are drawn at the effective wavelengths of the $gri$ bands.}
\label{fig:calerr}
\end{figure*}

The flux uncertainties from the spectroscopic calibration have strong correlations between spectral pixels at different wavelengths. This correlation mainly originates from the polynomial interpolation used in the calibration pipeline. In contrast to the independent statistical fluctuation due to the photon noise, the calibration uncertainties cause different pixels to vary in a correlated pattern. This makes it non-trivial to estimate the contribution of the calibration process to the Mg II flux uncertainty. 

We account for these correlated errors with an empirical method based on the tens of F-stars observed by the OzDES program for each field and epoch. Given that the F-stars do not vary intrinsically, the variations of the calibrated F-star spectra between epochs are due to the combination of independent statistical error (photon noise) and the calibration error. As an illustration, panel (a) of Figure \ref{fig:calerr} shows 24 spectroscopic epochs of the F-star DES J022526.89$-$044520.09. We calculate the mean spectrum $\langle f_{*,i} \rangle$, the rms spectrum $S_{*,i}$ and the mean statistical error $\sigma_{{\rm stat},*,i}$ as 
\begin{subequations}
\begin{align}
& \langle f_{*,i} \rangle = \sum_j f_{*,ij} / N_{\rm exp} \\
& S_{*,i} = \sqrt{\sum_j (f_{*,ij} - \langle f_{*,i} \rangle)^2 / (N_{\rm exp}-1)} \quad {\rm and} \\
& \sigma_{{\rm stat},*,i}^2 = \sum_j \sigma_{*,ij}^2 / N_{\rm exp}
\end{align}
\label{eq:fstar_rms}%
\end{subequations}
where $f_{*,ij}$ and $\sigma_{*,ij}^2$ are the flux and the statistical error of the $i^{\rm th}$ pixel in the $j^{\rm th}$ epoch and $N_{\rm exp}$ is the total number of epochs. Panel (b) of Figure \ref{fig:calerr} shows that the rms spectrum $S_{*,i}$ (black line) is broadly consistent with the estimated statistical error $\sigma_{{\rm stat},*,i}$ (red line). This indicates that the spectral variance is dominated by the statistical error at the pixel level, so we cannot directly extract the calibration uncertainty from the rms spectra. 

To reduce the statistical uncertainties, we bin the spectra with a bin size of 300 \AA.\, This averaging does not suppress the correlated uncertainties due to the calibration process. Panel (d) of Figure \ref{fig:calerr} shows the rms spectra of the binned F-star spectra in panel (c). The statistical errors become small relative to the spectral variations, so the binned rms spectra provide an estimate of only the calibration uncertainties. However, these calibration uncertainty estimates are still correlated and it is not straightforward to apply these to the quasar spectra. 

We adopt a Monte-Carlo method to determine the contribution of calibration uncertainties to the Mg II flux uncertainties. We first define a ``warping function''
\begin{equation}
W_{bj} = f_{*,bj} / \langle f_{*,b} \rangle
\label{eq:warping_func}
\end{equation}
where $f_{*,bj}$ is the flux of the $b^{\rm th}$ bin in the $j^{\rm th}$ epoch and $\langle f_{*,b} \rangle$ is the mean flux of the $b^{\rm th}$ bin over all epochs. We interpolate this discrete function with a 3rd order spline function to obtain a continuous warping function $W_j(\lambda)$. Each warping function $W_j(\lambda)$ represents one realization of the fractional spectral variation due to the correlated calibration uncertainties. Panel (e) of Figure \ref{fig:calerr} shows an example of the 24 warping functions corresponding to each epoch of the F-star spectra. The warping functions show the smallest variations around the centers of the photometric bands that anchor the calibrations (blue dotted lines). The variations are relatively large near the ends of the spectra, as well as around the location of the dichroic split between the red and blue arms of the spectrograph at $\sim$5700 \AA.
 
The next step is to multiply a quasar spectrum $f(\lambda)$ with a warping function $W_j(\lambda)$ to derive a warped realization of the spectrum
\begin{equation}
f_{{\rm warp},j}(\lambda) = f(\lambda) W_j(\lambda)
\label{eq:warped_spec}
\end{equation}
This represents a realization of the quasar spectra distorted by the calibration errors. We calculate the Mg II flux $F_{{\rm line,warp},j}$ from the warped spectra $f_{{\rm warp},j}(\lambda)$ as described in Section \ref{subsec:ironfit}. Each F-star epoch gives a warping function $W_j(\lambda)$ and therefore a Mg II flux realization $F_{{\rm line,warp},j}$. The standard deviation of the Mg II flux realizations after multiplying a spectrum by a set of warping functions is an estimate of the flux uncertainty due to the calibration. We convert this to the Mg II flux uncertainty of the observed spectra as 
\begin{equation}
\sigma_{\rm line,cal} = (\sigma_{\rm line,warp} / \langle F_{\rm line,warp} \rangle) \cdot F_{\rm line} 
\label{eq:calerr_Fline}
\end{equation}
where $\langle F_{\rm line,warp} \rangle$ and $\sigma_{\rm line,warp}$ are the mean and the standard deviation of the warped Mg II flux $F_{{\rm line,warp},j}$ over all realizations and $F_{\rm line}$ is the Mg II flux from the observed spectra. We add this calibration uncertainty $\sigma_{\rm line,cal}$ in quadrature with the Mg II flux uncertainty $\sigma_{\rm line}$. 

The warping functions are based on F-stars in the OzDES fields with at least five epochs, and we use multiple F-stars to create many warping functions. We exclude epochs where the median SNR of the F-star spectrum is less than 15. Panels (a) and (b) of Figure \ref{fig:calerr} show that there are occasionally bad pixels in the spectra that can affect the rms spectra. When binning the spectra, we remove a pixel if it is more than 4$\sigma$ away from the mean of the pixels at this wavelength over all epochs, where $\sigma$ is the flux uncertainty of this pixel. We remove bad bins where more than 10\% of the pixels are bad pixels, and exclude an epoch if more than 10\% of the bins are bad bins. We also manually exclude some epochs where there is contamination at $\sim$7100 \AA\, which originates from the LED in the 2dF gripper gantry. In summary, we create a total of 2116 warping functions based on 165 F-stars. 

\begin{figure}
\includegraphics[width=\linewidth]{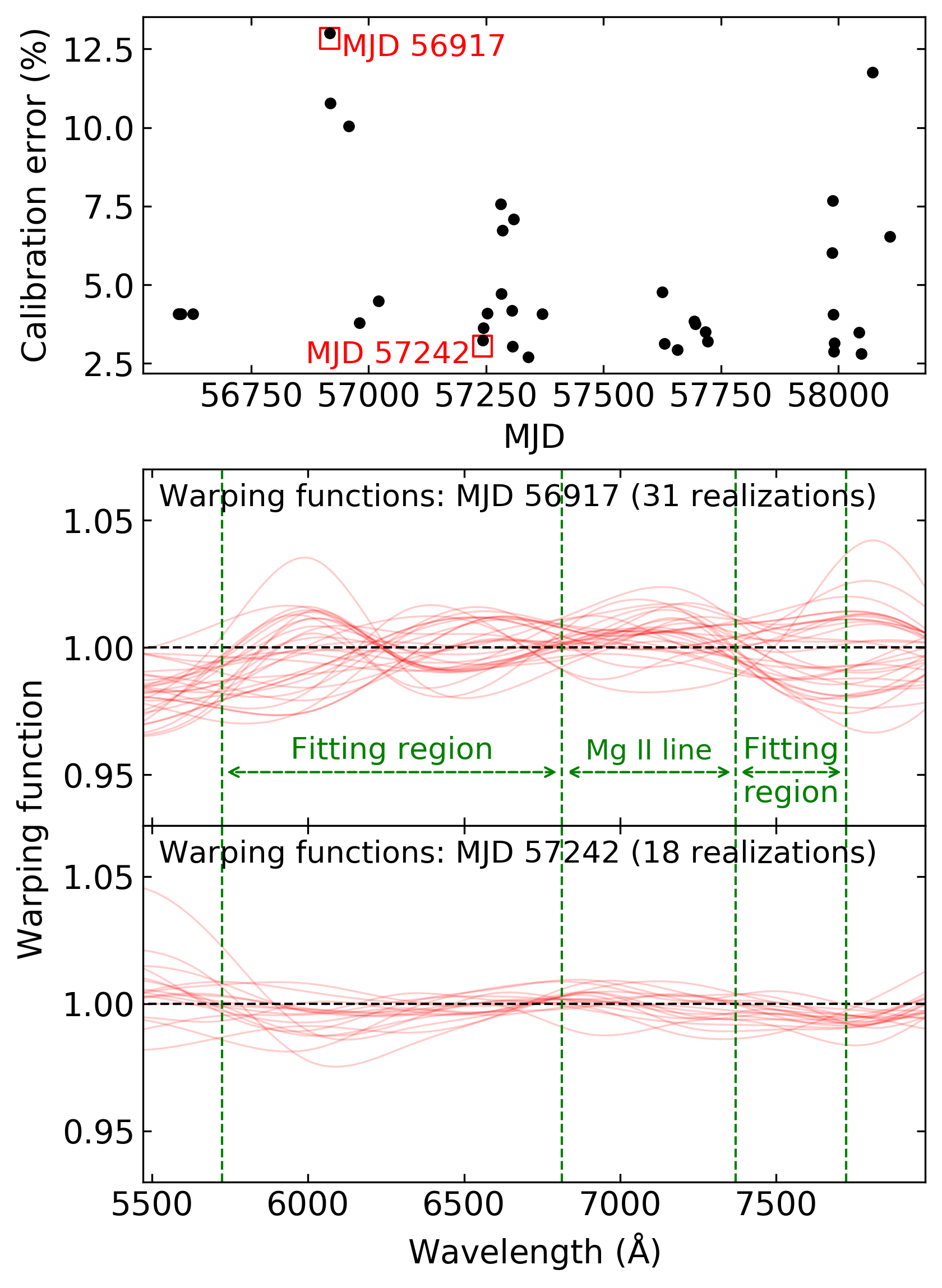}
\caption{Calibration errors and warping functions. ({\it Top panel}) Fractional calibration uncertainties of the Mg II line flux as a function of time for the quasar DES J003207.44$-$433049.00 at $z = 1.53$. ({\it Middle and bottom panels}) Warping functions corresponding to the epochs of MJD 56917 and MJD 57242 (red squares in the top panel) zoomed in around the Mg II line. The black dashed lines are drawn at unity. The green dashed lines mark the continuum $+$ iron fitting regions and the Mg II line region. Warping functions with larger variations correspond to larger calibration uncertainties.}
\label{fig:calerr_exp}
\end{figure}

For each epoch, there is effectively a library of warping functions based on the F-stars observed at that epoch. The variations in this library depend on the observational conditions for this epoch. We use this library to calculate the Mg II calibration uncertainties. The top panel of Figure \ref{fig:calerr_exp} shows the fractional calibration uncertainties of the Mg II line flux as a function of time for the quasar DES J003207.44$-$433049.00. The median calibration uncertainty is $\sim 4$\%, although it exceeds 10\% at some epochs. The middle and bottom panels show two examples of the warping functions near the Mg II line. The MJD 56917 epoch contains warping functions with large variations which correspond to a large calibration uncertainty, while the small variations in MJD 57242 imply a small calibration uncertainty. The calibration uncertainty contributes more than half of the Mg II flux uncertainty for three sources in our main sample, while the photon noise dominates the error budget for the other sources. 

%Time Series Analysis 
\section{Time Series Analysis} \label{sec:lcanl}

We create continuum lightcurves in the $g$ band from the DES database and create Mg II line lightcurves using the methodology of Section \ref{sec:specanl}. We adopt the V01 template as the iron template for our standard analyses. The other two templates either use a similar method
to the V01 or directly use the V01 template except for the Mg II line region. We examine the effect of using the other two templates on the lag measurements. We visually inspect and remove epochs where the continuum $+$ iron model fails to match the spectrum or where the spectrum is contaminated by known instrumental artifacts like the 7100 \AA\, bump due to scattered LED light. The lightcurves of our main sample are available in the online journal.

\begin{table*}
\small
\renewcommand{\arraystretch}{1.25}
\begin{tabular}{ccllcccc}
\hline
Source Name & z & JAVELIN Lag & ICCF Lag & FPR & flag & $\sigma_{\rm line}$ & log($M_{\rm BH}/M_{\odot}$)\\
 &  & (days) & (days) &  &  & (km/s) & \\
\hline
DES J025225.52$+$003405.90 & 1.62 & $521.7^{+44.5}_{-49.9}$ & $465.1^{+39.5}_{-33.8}$ & 0.068 & 0 & 2654.7 & 9.07 \\
DES J021612.83$-$044634.10 & 1.56 & $131.8^{+36.8}_{-22.9}$ & $141.6^{+28.3}_{-40.1}$ & 0.037 & 0 & 1727.2 & 8.11 \\
DES J033553.51$-$275044.70 & 1.58 & $124.1^{+56.8}_{-22.7}$ & $146.4^{+31.8}_{-31.6}$ & 0.000 & 0 & 2719.6 & 8.48 \\
DES J003710.86$-$444048.11 & 1.07 & $396.4^{+57.1}_{-38.2}$ & $458.4^{+27.8}_{-32.8}$ & 0.062 & 0 & 2399.2 & 8.97 \\
DES J003207.44$-$433049.00 & 1.53 & $372.2^{+6.1}_{-2.2}$ & $368.7^{+22.1}_{-32.4}$ & 0.039 & 0 & 1904.4 & 8.65 \\
DES J003015.00$-$430333.52 & 1.65 & $491.7^{+38.6}_{-12.5}$ & $495.6^{+28.6}_{-34.0}$ & 0.056 & 0 & 3924.3 & 9.38 \\
DES J003052.76$-$430301.08 & 1.43 & $404.8^{+26.7}_{-26.4}$ & $395.3^{+33.0}_{-29.7}$ & 0.154 & 1 & 2101.0 & 8.79 \\
DES J003234.33$-$431937.81 & 1.64 & $657.0^{+47.8}_{-30.7}$ & $654.6^{+14.1}_{-18.5}$ & 0.000 & 0 & 1805.0 & 8.83 \\
DES J003206.50$-$425325.22 & 1.75 & $433.9^{+35.1}_{-13.6}$ & $456.6^{+41.2}_{-42.5}$ & 0.029 & 0 & 3767.8 & 9.28 \\
\hline
\end{tabular}
\caption{Column (1) gives the name of the sources. Columns (2) gives the redshifts. Column (3) and (4) give the observed-frame lags and uncertainties from JAVELIN and the ICCF method, respectively. Column (5) gives the FPR from simulations. Column (6) gives the flags. ${\rm flag}=1$ means that the source has a FPR larger than 0.1 (Section \ref{sec:simulations}). Column (7) gives the line dispersion. Column (8) gives the black hole mass. The uncertainty of the black hole mass is about 0.4 dex.}
\label{tab:result}
\end{table*}

%Lag measurement
\subsection{Lag Measurement} \label{subsec:lagfit}
We measure the lags with JAVELIN \citep[e.g.,][]{Zu2011,Zu2013} and the interpolated cross-correlation function \citep[ICCF, e.g.,][]{Gaskell1987,Peterson1998,Peterson2004}. JAVELIN models the quasar variability with a damped random walk (DRW) and assumes that the line lightcurve is a shifted, scaled and top-hat smoothed version of the continuum lightcurve. It first fits the continuum lightcurve using a MCMC sampler to constrain the DRW amplitude and characteristic time scale. With these constraints as the prior, it then fits the continuum and the line lightcurves simultaneously to derive the posterior distributions of the transfer function width, the scaling factor, and the time lag. We set a lag limit of [$-$1000,1000] days and allow the DRW amplitude and time scale, the transfer function width and the scaling factor to vary freely during the fitting. While our data cannot well constrain the DRW time scale, previous studies found that it does not have significant impact on the lag measurements \citep[e.g.,][]{Yu2020_RMErr,Homayouni2020}.

The ICCF method calculates the cross-correlation function (CCF) after linearly interpolating the lightcurves. It uses the centroid or the peak of the CCF as the lag estimate. To estimate the lag uncertainty, it randomizes the lightcurve according to the single-epoch uncertainties and/or randomly selects a sub-sample of the epochs with replacement to create a number of lightcurve realizations. The center or the peak of the CCF for each realization forms the cross-correlation center distribution (CCCD) or the cross-correlation peak distribution (CCPD). The scatter of the CCCD or the CCPD gives an estimate of the lag uncertainty. We use the python implementation \textsf{PyCCF} \citep{pyccf} for the ICCF method. We create 8000 realizations with both flux randomization and the random sub-sampling and use realizations with $r_{\rm peak}>0.5$ to estimate the lag uncertainties, where $r_{\rm peak}$ is the peak cross-correlation coefficient of the CCF. For each realization, we adopt the region with $r>0.8r_{\rm peak}$ to calculate the CCF center or peak. Previous studies found that the CCCD generally yields better lag measurements \citep[e.g.,][]{Peterson2004}, so we adopt the CCCD as the reference lag distribution of the ICCF method.

%Lag selection
\subsection{Aliasing Removal \& Lag Selections} \label{subsec:lagselection}
Secondary peaks in the lag distributions due to aliasing can bias the lag measurements. We adopt the aliasing removal algorithm from the SDSS RM project \citep[e.g.,][]{Grier2019,Homayouni2020}. We first weight the lag distributions with a convolution of two weighting functions. The first weighting function is
\begin{equation}
P(\tau) = [N(\tau)/N(0)]^2
\label{eq:weight_sdss}
\end{equation}
where $N(\tau)$ is the number of overlapping epochs between the line lightcurve and the continuum lightcurve shifted by the lag $\tau$. This decreases the likelihood that we will measure a lag when the line lightcurve has little overlap with the continuum after it is shifted by the lag. We force $P(\tau)$ to be symmetric by calculating only $P(\tau>0)$ and setting $P(-\tau) = P(\tau)$. The second weighting function is the auto-correlation function (ACF) of the continuum lightcurve. This accounts for how rapidly the continuum varies. We set the ACF to zero when it drops below zero. 

We then smooth the weighted lag distributions with a Gaussian kernel with a width of 12 days. We identify the highest peak in the weighted, smoothed lag distribution as the major peak. We define the best-fit lag as the median of the unweighted lag distributions within the major peak and calculate the lag uncertainty using the 16th and 84th percentiles. The identification of the major peak region can be ambiguous when there are multiple connected peaks. We define a peak to be separate from other peaks if it is more than 10 days away from the neighboring peaks and its prominence is larger than 10\% of the neighboring peaks. We define the peak prominence based on the python module \textsf{scipy.signal.peak\_prominences} \citep{scipy}. It extends a horizontal line from the peak until it intersects with a higher peak or the lag boundaries. On each side of the peak, it searches for a minimum between the peak and the intersection. It calculates the peak prominence as the vertical distance between the peak and the higher minimum on the two sides.   

We define a lag as significant if (1) $f_{\rm peak} > 0.6$, where $f_{\rm peak}$ is the fraction of the total probability included in the major peak and (2) the lag is above or below zero with at least $3\sigma$ significance. We consider a lag measurement for a quasar successful if both the JAVELIN and ICCF methods yield significant lags and the two algorithms are consistent within $2\sigma$. We obtain successful lag measurements for 9 sources. All lag measurements are positive -- no negative lags pass the selection criteria. We refer to this as the ``main sample''. 

\begin{figure*}
\includegraphics[width=0.93\linewidth]{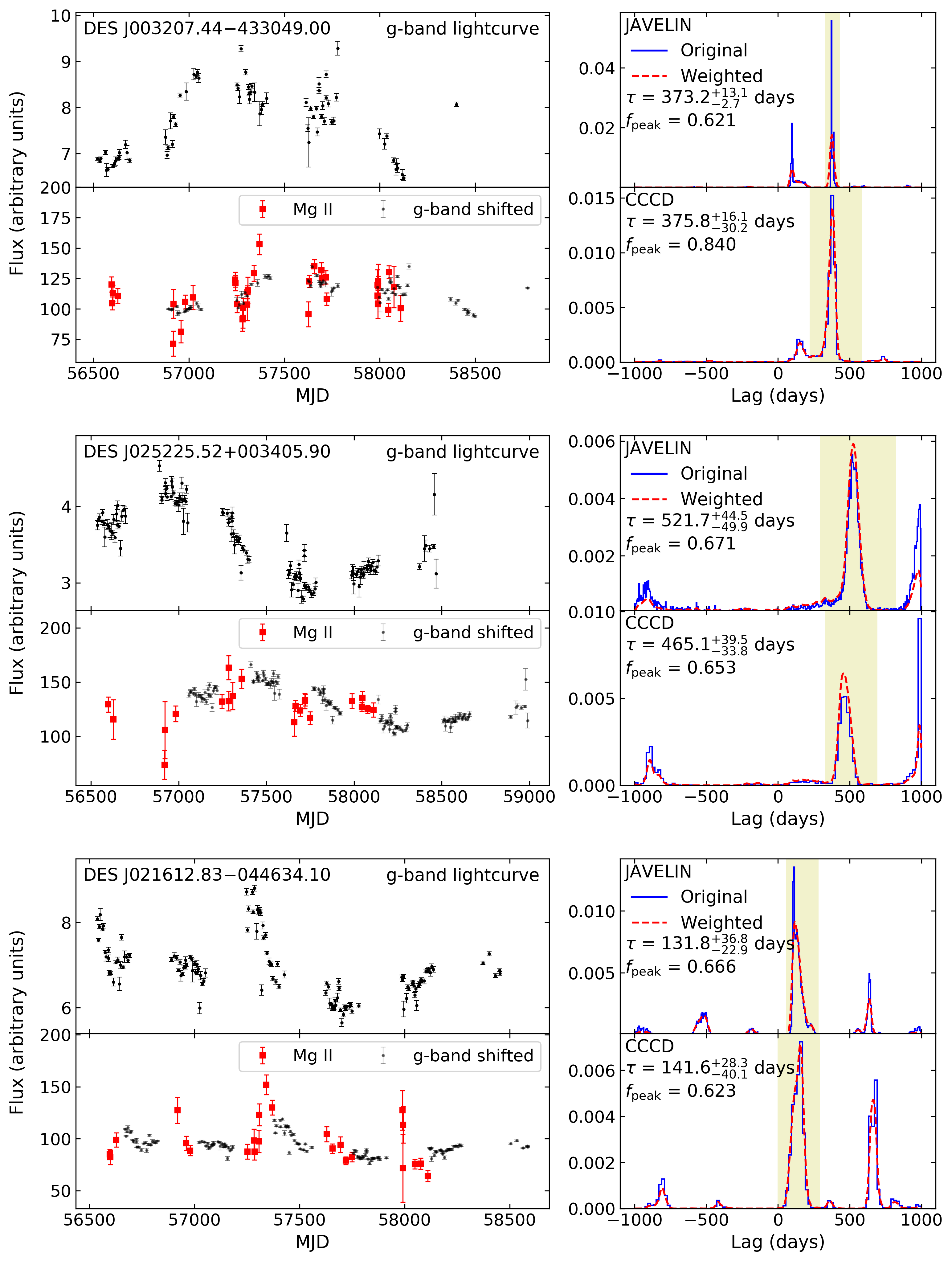}
\caption{Lightcurves and lag measurements. Each 2$\times$2 grid of panels presents a source with the name shown in the upper left corner. ({\it Left column}) $g$-band continuum lightcurve and Mg II line lightcurve. The black points in the upper panel and the red squares in the lower panel represent the continuum and line lightcurve, respectively. The black points in the lower panel represent the continuum lightcurve shifted by the best-fit JAVELIN lag. We re-scale the shifted continuum lightcurve so that its median matches the median of the line lightcurve. ({\it Right column}) JAVELIN and ICCF lag distributions. The blue solid lines in the upper and lower panels show the JAVELIN posterior lag distribution and the CCCD from the ICCF method, respectively. The red dashed lines represent the weighted and smoothed lag distributions. The yellow shaded regions represent the ``major peak'' ranges.}
\label{fig:lag_gp0}
\end{figure*}

\begin{figure*}
\includegraphics[width=\linewidth]{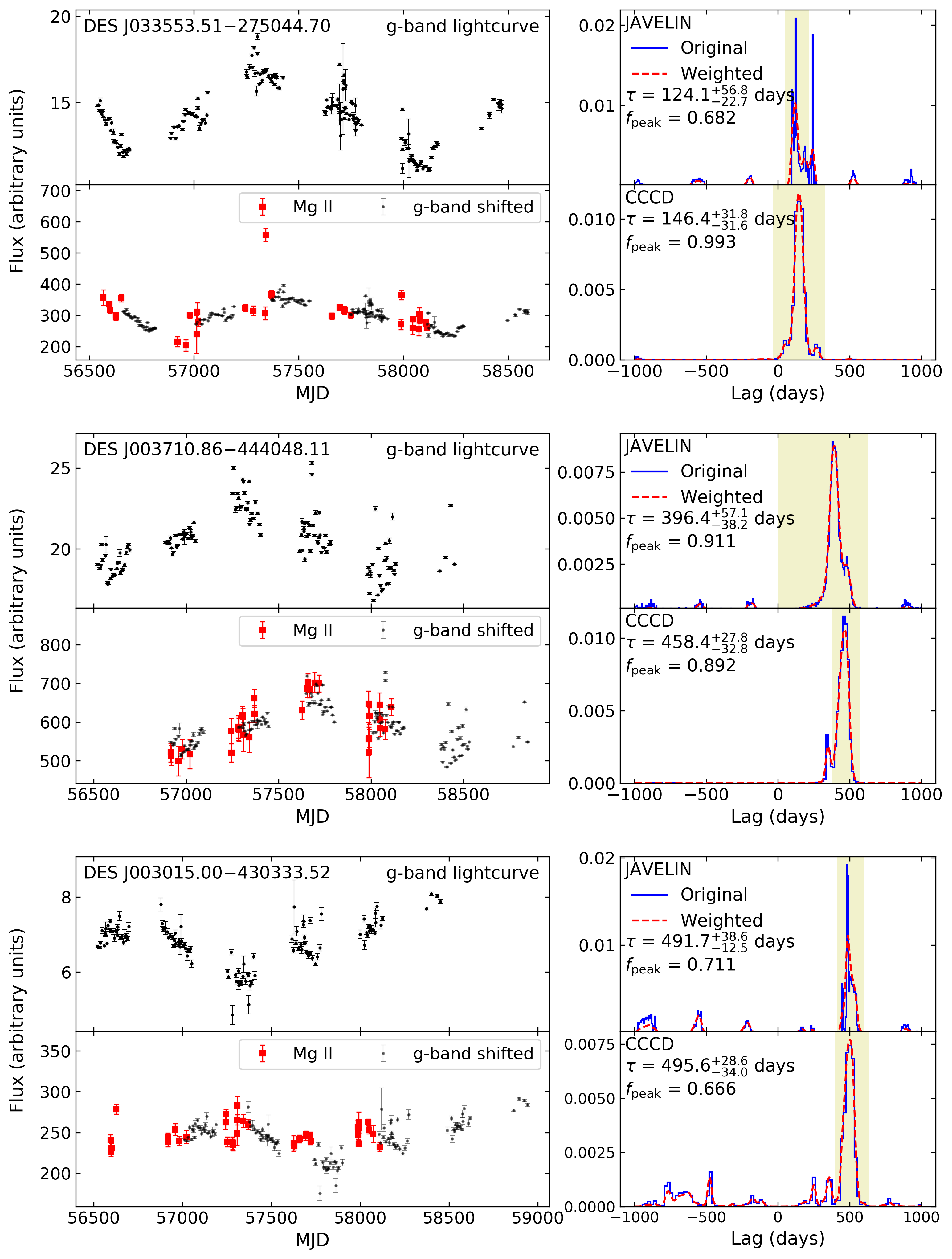}
\caption{Figure \ref{fig:lag_gp0}, continued.}
\label{fig:lag_gp1}
\end{figure*}

\begin{figure*}
\includegraphics[width=\linewidth]{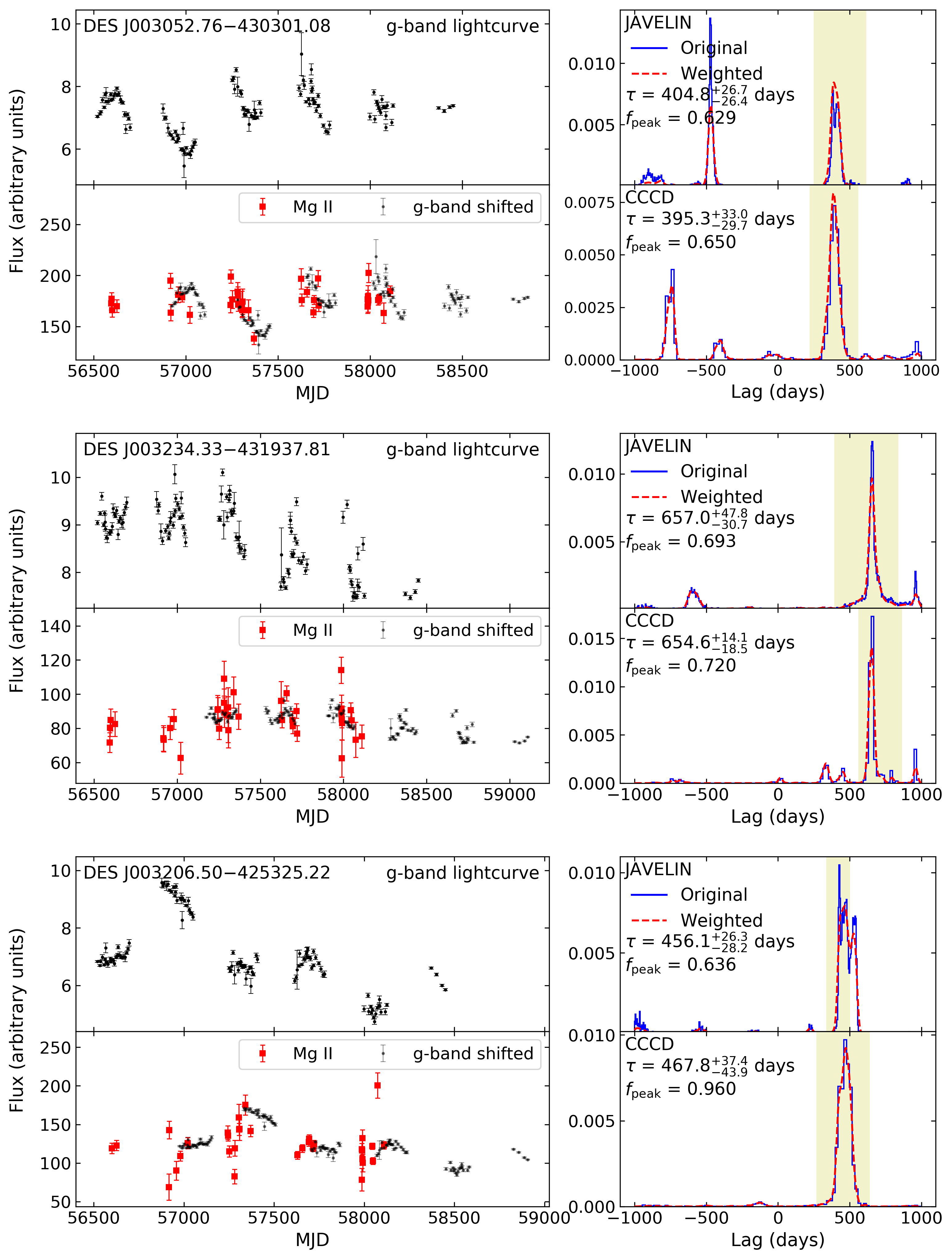}
\caption{Figure \ref{fig:lag_gp0}, continued.}
\label{fig:lag_gp2}
\end{figure*}

Table \ref{tab:result} provides the lags for the main sample. Figures \ref{fig:lag_gp0} - \ref{fig:lag_gp2} show the lightcurves (left columns) and the lag distributions (right columns). We shift the continuum lightcurves by the best-fit JAVELIN lags to compare with the line lightcurves. We re-scale each shifted continuum lightcurve so that its median matches the median of the line lightcurve. For about half of the main sample sources, the shifted continuum lightcurve overlaps well with the line lightcurve. This overlap further supports the reliability of the lag. The shifted continuum lightcurves of the other sources have less overlap with the line lightcurves, but they form a reasonable ``interpolation'' for the line lightcurves between the neighboring seasons. All lag distributions show clear major peaks that agree between the JAVELIN and the ICCF results. The sources that have good overlap between the shifted continuum and the line lightcurves tend to have lags with smaller uncertainties. An extension of the observational seasons would be very helpful for lag measurements in future RM campaigns. As previous studies found that JAVELIN generally produces more robust lag determinations with more realistic uncertainties \citep[e.g.,][]{Li2019,Yu2020_RMErr}, we adopt the JAVELIN lags as the reference lags for the remainder of our analyses. 

\begin{figure*}
\includegraphics[width=0.95\linewidth]{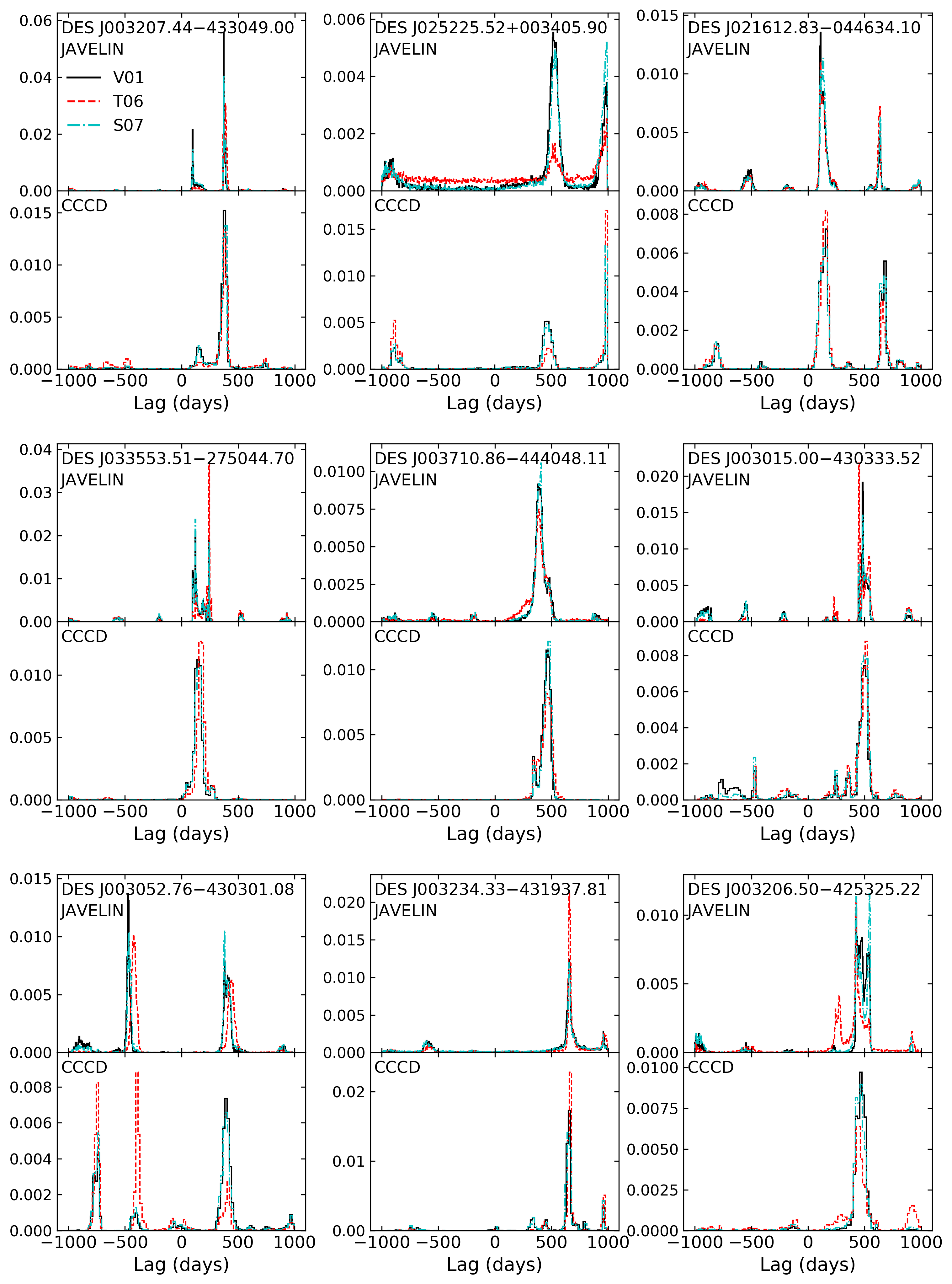}
\caption{Comparison of lag distributions for the three different iron templates. Each pair of panels show the JAVELIN (upper panel) and the ICCF (lower panel) lag distributions for a source with the name shown in the upper left corner. The black solid, red dashed and cyan dash-dotted lines represent the results from the V01, T06 and S07 templates, respectively.}
\label{fig:lag_tplcomp_gp0}
\end{figure*}

While we use V01 as our standard iron template, we compare the lags derived using the other two iron templates in Figure \ref{fig:lag_tplcomp_gp0}. The lag distributions from the three iron templates agree well. In some cases, such as the JAVELIN lags for DES J025225.52+003405.90 and the ICCF lags for DES J003052.76-430301.08, the relative strength of the peaks can differ between templates. For the T06 results of DES J025225.52+003405.90, the major peak identification remains the same as the other two templates and the lag measurements are similar. For DES J003052.76-430301.08, the stronger aliasing in the T06 results can lead to a mis-identification of the major peak at negative lags. This source is also a flagged source that we discuss further in Section \ref{sec:simulations}. Overall, the choice of iron template does not have a significant impact on our lag measurements. 


%Simulations 
\section{Lag Reliability} \label{sec:simulations}

\begin{figure}
\includegraphics[width=\linewidth]{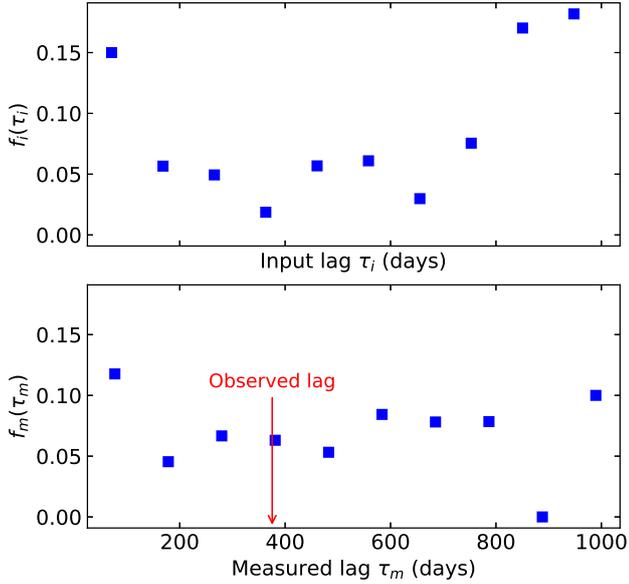}
\caption{False Positive Rate (FPR) as a function of the input lag ({\it upper panel}) and the measured lag ({\it lower panel}) in the simulations for DES J003207.44$-$433049.00. The blue squares show the FPR for the ten lag bins. The FPR is defined as the number of false positive realizations divided by the total number of realizations that pass the lag selection criteria in each bin. The red arrow in the lower panel marks the lag from the observed lightcurve.}
\label{fig:FPR_examp}
\end{figure}

\begin{figure}
\includegraphics[width=\linewidth]{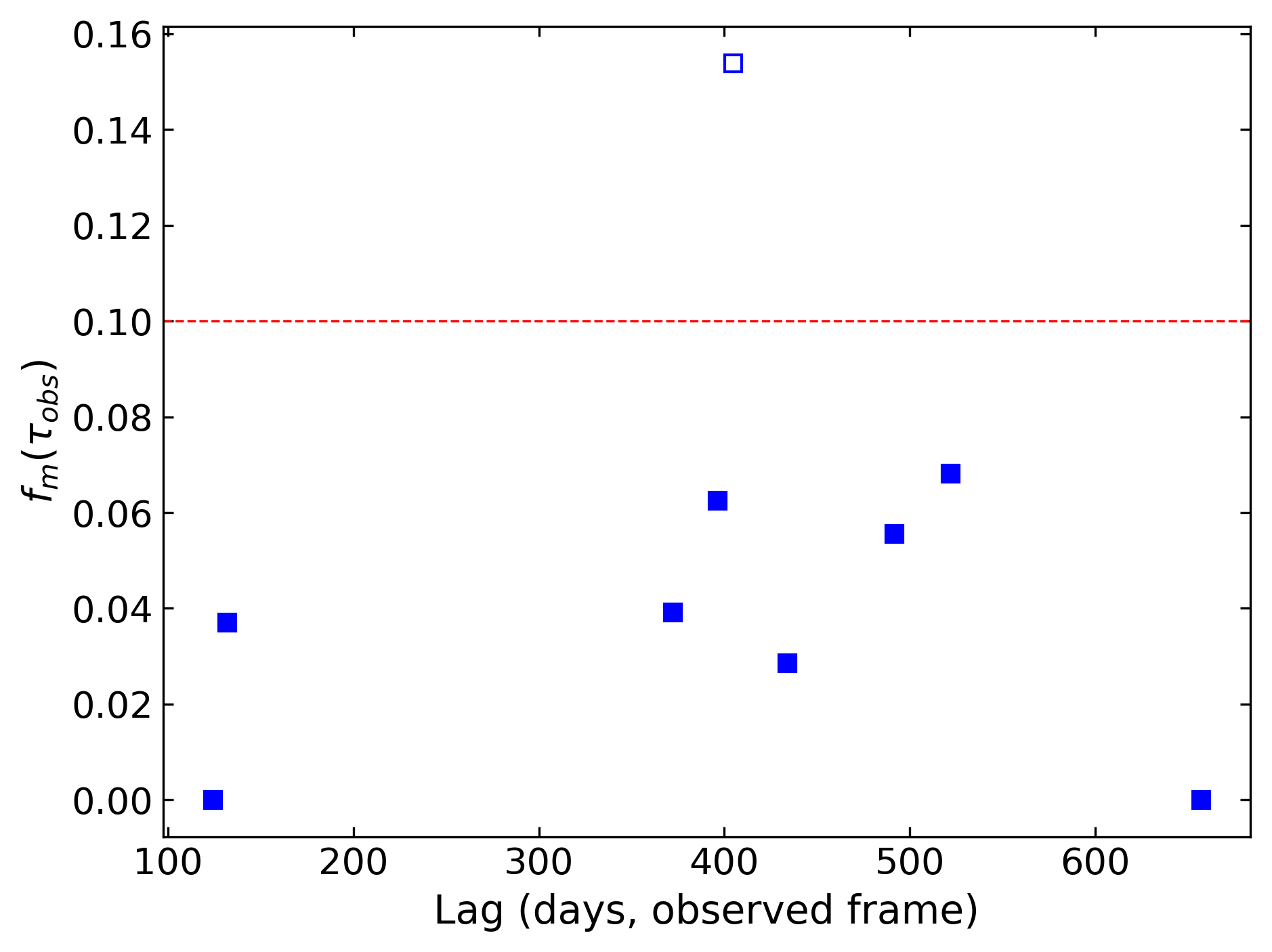}
\caption{False Positive Rate (FPR) of the main sample. Each square represents the FPR of one source in the main sample at the observed lag. The red dashed line is drawn at ${\rm FPR} = 0.1$. The filled and empty symbols represent the sources with the FPR below or above the cut, respectively.}
\label{fig:FPR_obs}
\end{figure}

We assess the reliability of the lags using the simulation tool presented by \citet{Penton2021}. We simulate the continuum for each source with 2000 DRW lightcurve realizations, where the continuum variability is matched to the variability of the observed lightcurves. For each realization, we create the simulated line lightcurve using a top-hat transfer function with an input lag $\tau_{\rm i}$ randomly drawn between 10 days and 1000 days. The simulated lightcurves are produced with the same cadence and SNR as the observed lightcurves. We perform the same analysis on the simulated lightcurve pairs as we did for the observed lightcurves in Section \ref{sec:lcanl}. For the $N_{\rm p}$ realizations that pass the lag selection criteria, we compare the measured lag $\tau_{\rm m}$ with the input lag $\tau_{\rm i}$. We define the number of false positive realizations $N_{\rm bad}$ as those where the measured lag $\tau_{\rm m}$ and the input lag $\tau_{\rm i}$ differ by 3$\sigma$.

The number of the false positives has a distribution function $N_{\rm bad,i} (\tau_{\rm i})$ over the input lag $\tau_{\rm i}$ or a distribution function $N_{\rm bad,m} (\tau_{\rm m})$ over the measured lag $\tau_{\rm m}$. These two distribution functions can have different shapes. Similarly, the number of realizations that pass the selection cuts has two distribution functions $N_{\rm p,i} (\tau_{\rm i})$ and $N_{\rm p,m} (\tau_{\rm m})$. We define two false positive rates (FPR)
\begin{subequations}
\begin{align}
& f_{\rm i}(\tau_{\rm i}) = N_{\rm bad,i} (\tau_{\rm i}) / N_{\rm p,i} (\tau_{\rm i}) \quad {\rm and}\\
& f_{\rm m}(\tau_{\rm m}) = N_{\rm bad,m} (\tau_{\rm m}) / N_{\rm p,m} (\tau_{\rm m})
\end{align}
\label{eq:FPR}%
\end{subequations}
The first expression $f_{\rm i}(\tau_{\rm i})$ is our estimate of the probability that we measure an incorrect lag for a source with a true lag $\tau_{\rm i}$, while $f_{\rm m}(\tau_{\rm m})$ is our estimate of the probability that the measured lag is wrong when we measure a lag $\tau_{\rm m}$. 

Figure \ref{fig:FPR_examp} shows an example of $f_{\rm i}(\tau_{\rm i})$ and $f_{\rm m}(\tau_{\rm m})$. We divide the lags into 10 bins and calculate the FPR in each bin using Equation \ref{eq:FPR}. The FPR function $f_{\rm i}(\tau_{\rm i})$ shows local minima around one year and two years, while it increases toward the two boundaries. This indicates that our data are most sensitive to lags where the shifted continuum overlaps well with the line lightcurves, while it is more difficult to correctly recover lags that are extremely small or large, or where the lags are around the seasonal gaps. The function $f_{\rm m}(\tau_{\rm m})$ also increases toward the two boundaries, while the trend is flatter than $f_{\rm i}(\tau_{\rm i})$. The FPR functions $f_{\rm i}(\tau_{\rm i})$ and $f_{\rm m}(\tau_{\rm m})$ in Figure \ref{fig:FPR_examp} were calculated with the same observational cadence and SNR. The cadence and SNR can vary among different sources, so sources where $\tau_{\rm i}$ or $\tau_{\rm m}$ falls into the seasonal gaps do not necessarily have larger FPR.

Given that we do not know the true lags, the definition $f_{\rm m}(\tau_{\rm m})$ is more suitable for assessing the reliability of the measured lags. We define a fiducial FPR for each source as
\begin{equation}
{\rm FPR} = f_{\rm m}(\tau_{\rm obs}) = N_{\rm bad,m} (\tau_{\rm obs}) / N_{\rm p,m} (\tau_{\rm obs})
\label{eq:FPR_obs}
\end{equation}
where $\tau_{\rm obs}$ is the lag measured from the observed lightcurves. Given the finite number of realizations with discrete measured lags, we define $N_{\rm p,m} (\tau_{\rm obs})$ as the number of realizations where the measured lag $\tau_{\rm m}$ agrees with the observed lag $\tau_{\rm obs}$ to within $1\sigma$. Table \ref{tab:result} gives the FPR for the main sample sources and Figure \ref{fig:FPR_obs} shows the FPR against the observed lags. The median FPR of the main sample is about 4\%. We set a threshold at ${\rm FPR} = 0.1$ and add ${\rm flag} = 1$ to the one source above the threshold in Table \ref{tab:result}. This flag does not mean the source is a spurious detection. A FPR of $\sim 15\%$ is still statistically small and the flag is just to separate this source from others that have significantly lower FPRs.

The simulation method has been commonly used to assess the lag reliability \citep[e.g.,][]{Shen2015,Li2019,Yu2020_RMErr,Penton2021}, although these simulations have different configurations, such as the choice of the input lags and how the observations constrain the simulated lightcurves. This method differs from \citet{Homayouni2020} in that they use uncorrelated lightcurves to estimate the FPR. The two different methods address two different aspects of the lag reliability. The \citet{Homayouni2020} method estimates the probability that one has measured a lag from a lightcurve where there is no intrinsic lag signal. In contrast, our method first determines if there is a real lag signal based on strict lag selection criteria, and then uses simulations to quantify the probability that the measured lag is the true lag.

%R-L relation, BH mass
\section{Black Hole Mass and R - L Relation} \label{sec:RL_BHMass}

\begin{figure*}
\includegraphics[width=0.95\linewidth]{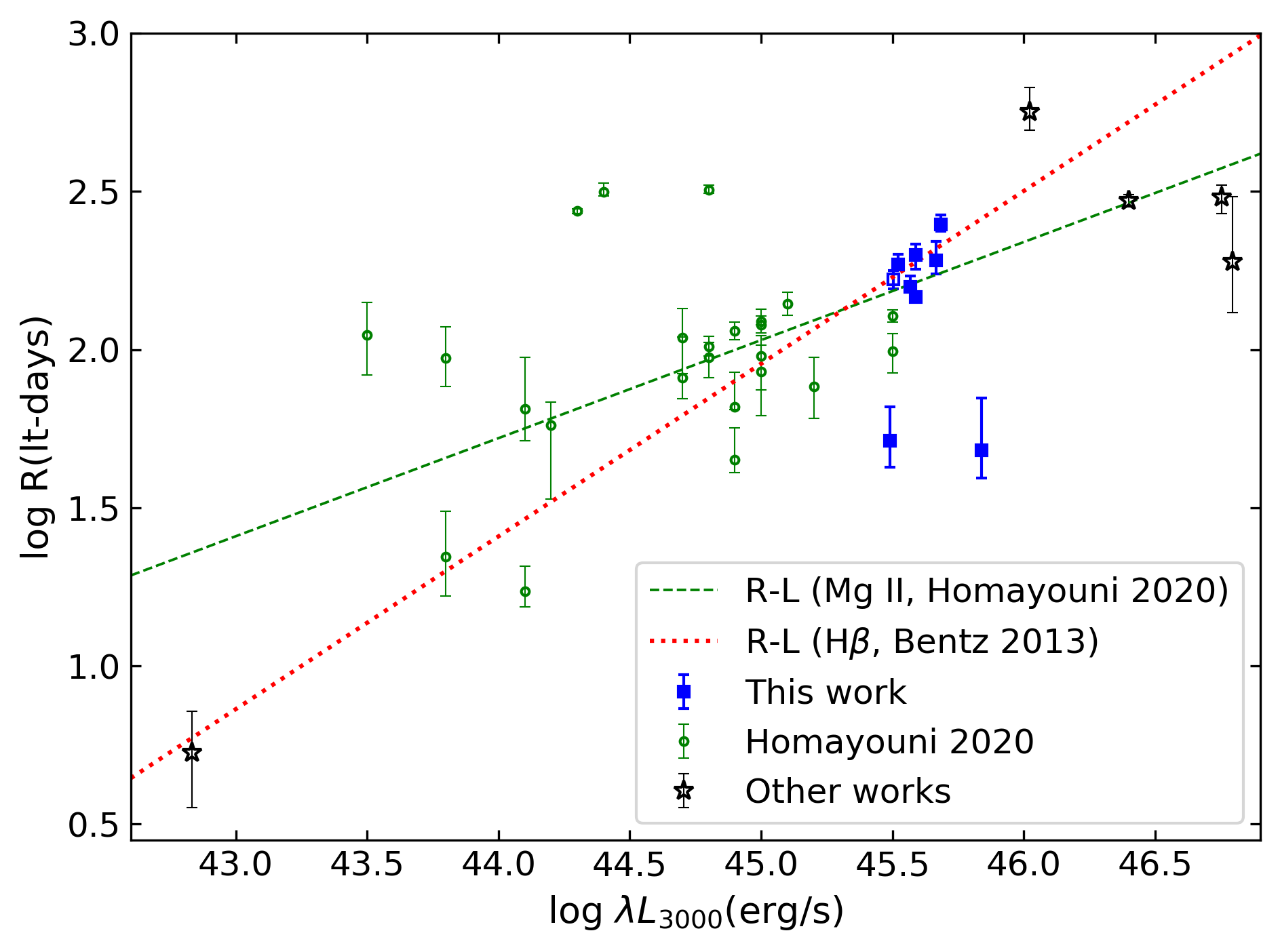}
\caption{Broad line region radius $-$ luminosity ($R - L$) relation based on Mg II lag measurements. The blue squares show our measurements. The filled and empty symbols represent the clean and flagged sources, respectively. The green circles show the ``gold sample'' of \citet{Homayouni2020} and the green dashed line shows their best-fit Mg II $R - L$ relation with a slope of $\sim0.3$. The black points show the measurements from \citet{Metzroth2006}, \citet{Lira2018}, \citet{Czerny2019}, \citet{Zajacek2020} and \citet{Zajacek2021}. The red dotted line is the H$\beta$ $R - L$ relation by \citet{Bentz2013} after we convert the monochromatic luminosity from 3000 \AA\, to 5100 \AA\ with the bolometric corrections from \citet{Richards2006}. The H$\beta$ $R - L$ relation has a slope of $\sim0.5$.}
\label{fig:RL}
\end{figure*}

We use the line dispersion to parametrize the Mg II line width from the co-added spectra after subtracting the continuum and the iron emission. Given the line profile $P(\lambda)$, the line dispersion is 
\begin{equation}
\sigma_{\rm line}^2 = \left[\int \lambda^2 P(\lambda) d\lambda \, \Big/ \int P(\lambda) d\lambda \right] - P_0(\lambda)^2
\label{eq:linesig}
\end{equation}
where $P_0(\lambda)$ is the first moment of the line profile. Previous studies have shown that the line dispersion is a better line width estimator for the black hole mass than the full-width half maximum \citep[e.g.,][]{Peterson2004,Dalla2020}. The S/N of our data is too low to provide reliable line width measurements from the rms spectra that are commonly used in RM studies. However, the line dispersion from the mean spectra is generally a reasonable proxy to that from the rms spectra \citep[e.g.,][]{Wang2020,Dalla2020}.

We calculate black hole masses using Equation \ref{eq:bhmass} with the measured lag and the line dispersion and adopt the virial factor $f=4.31$ from \citet{Grier2013Martini}. Table \ref{tab:result} lists the line dispersions and the black hole masses for the main sample. The uncertainty of the black hole mass is about 0.4 dex, mainly due to the intrinsic scatter in calibrating the virial factor \citep[e.g.,][]{Peterson2014}. Other measurement errors are generally small relative to that 0.4-dex uncertainty, so we do not consider more detailed uncertainty estimates for the individual black hole masses. 

We compare our lags to the $R - L$ relations from previous studies in Figure \ref{fig:RL}. Our sample covers a luminosity range where there are few existing Mg II lag measurements. The Mg II lags tightly fall on the H$\beta$ $R - L$ relation by \citet{Bentz2013} that has a slope of $\sim0.5$ with the exception of the two outliers, in contrast to the large intrinsic scatter found by \citet{Homayouni2020}. The consistency between the observed Mg II lags and the predictions of the H$\beta$ $R - L$ relation may indicate that the Mg II emitting radius is similar to the H$\beta$ radius. However, the predicted lags by the \citet{Bentz2013} and \citet{Homayouni2020} $R - L$ relations are similar within the luminosity range of our sample, so our results do not rule out the \citet{Homayouni2020} $R - L$ relation that has a shallower slope than \citet{Bentz2013}.

The Mg II line is dominated by collisional excitation, in contrast to the recombination dominated Balmer lines. Theoretical studies have predicted that the average emitting radius for the Mg II should be larger than for the Balmer lines \citep[e.g.,][]{Korista2000,Guo2020}. Observational studies, on the other hand, have found that the Mg II lags are similar to the H$\beta$ lags for the same source, although only $\sim10$ sources have lag measurements for both Mg II and H$\beta$ \citep[e.g.,][]{Clavel1991,Metzroth2006,Homayouni2020}. The consistency between the H$\beta$ $R - L$ relation and the measured Mg II lags in our sample is in general agreement with the previous observational results. 

We calculate the bolometric luminosity using the monochromatic luminosity at 3000 \AA\, and the bolometric corrections by \citet{Richards2006}. We divide the bolometric luminosity by the Eddington luminosity to calculate the Eddington ratio. We get Eddington ratios $\sim 1$ for the two sources that have significantly smaller lags than the prediction of the $R - L$ relation. This generally agrees with previous studies that found smaller lags from sources with larger Eddington ratios \citep[e.g.,][]{Du2016,Du2018,Dalla2020,Martinez2020}. However, the Eddington ratio depends on the black hole mass and therefore the reliability of the lag measurements. Verification of the correlation between lags and Eddington ratios requires independent black hole mass estimates, which is beyond the scope of this paper.

%Summary
\section{Summary} \label{sec:summary}

We have measured robust Mg II lags for nine quasars that were observed as part of the OzDES RM program. These quasars have both photometric observations from DES and spectroscopic observations from OzDES, where the DES photometry provides both well-sampled continuum lightcurves and a means to calibrate the spectroscopic data based on the calibration pipeline developed by \citet{Hoormann2019}. The Mg II region is severely affected by iron emission, especially many Fe II multiplets, and we have developed a MCMC-based algorithm to account for the continuum $+$ iron emission near the Mg II line. We model the continuum with a power-law and model the iron emission with smoothed iron templates from the literature. A second complication is that there are significant, correlated errors in these multi-fiber spectra due to the calibration process. We have developed a new method to use multi-epoch spectra of F-stars to characterize the correlated errors from the spectroscopic calibrations. The main results are as follows:

\begin{enumerate}
\item We measure nine positive Mg II lags and no negative lags. The different iron templates have little effect on the lag measurements. We use simulations to assess the lag reliability and obtain a median FPR of about 4\%. The FPR characterizes the probability that the measured lag is inconsistent with the true lag.
\\
\item After accounting for the difference in the typical continuum near H$\beta$ and Mg II, the Mg II lag measurements are consistent with both the H$\beta$ $R - L$ relation by \citet{Bentz2013} and the Mg II $R - L$ relation by \citet{Homayouni2020} with the exception of the two outliers. The agreement with the H$\beta$ $R - L$ relation suggests that Mg II is emitted at a similar radius to H$\beta$, in general agreement with previous observational results \citep[e.g.,][]{Clavel1991,Metzroth2006,Homayouni2020}.
\end{enumerate}

In this paper we only consider luminosity ranges where the continuum lightcurves overlap well with the line lightcurves after they have been shifted by the expected lags. While our sample fills a luminosity range that had few robust measurements, the dynamic range of our sample is too small to independently derive a robust estimate of both the $R - L$ relation and its intrinsic scatter. Furthermore, previous studies are not homogeneous in their approach to lag measurements, uncertainty estimates, and lag quality cuts. Simply combining all the existing results may lead to a biased estimate of the $R - L$ relation. We therefore do not fit a new $R - L$ relation. In our next paper we will expand our analysis to the full OzDES RM sample with Mg II data, take into account how systematic errors impact the measurement of the $R-L$ relation, and conduct a homogeneous re-analysis of existing results with sufficient public information.

\section*{Acknowledgements}

We thank the anonymous referee for the careful review and comments. We thank Dr. Yue Shen for the constructive comments. ZY was supported in part by the United States National Science Foundation under Grant No. 161553 to PM. PM is grateful for support from the Radcliffe Institute for Advanced Study at Harvard University. PM also acknowledges support from the United States Department of Energy, Office of High Energy Physics under Award Number DE-SC-0011726.

AP and UM are supported by the Australian Government Research Training Program (RTP) Scholarship.

This research was supported in part by the Australian Government through the Australian Research Council Laureate Fellowship funding scheme (project FL180100168). 

Based in part on data acquired at the Anglo-Australian Telescope, under program A/2013B/012]. We acknowledge the traditional owners of the land on which the AAT stands, the Gamilaroi people, and pay our respects to elders past and present.

Funding for the DES Projects has been provided by the U.S. Department of Energy, the U.S. National Science Foundation, the Ministry of Science and Education of Spain, the Science and Technology Facilities Council of the United Kingdom, the Higher Education Funding Council for England, the National Center for Supercomputing Applications at the University of Illinois at Urbana-Champaign, the Kavli Institute of Cosmological Physics at the University of Chicago, the Center for Cosmology and Astro-Particle Physics at the Ohio State University, the Mitchell Institute for Fundamental Physics and Astronomy at Texas A\&M University, Financiadora de Estudos e Projetos, Funda{\c c}{\~a}o Carlos Chagas Filho de Amparo {\`a} Pesquisa do Estado do Rio de Janeiro, Conselho Nacional de Desenvolvimento Cient{\'i}fico e Tecnol{\'o}gico and the Minist{\'e}rio da Ci{\^e}ncia, Tecnologia e Inova{\c c}{\~a}o, the Deutsche Forschungsgemeinschaft and the Collaborating Institutions in the Dark Energy Survey. 

The Collaborating Institutions are Argonne National Laboratory, the University of California at Santa Cruz, the University of Cambridge, Centro de Investigaciones Energ{\'e}ticas, 
Medioambientales y Tecnol{\'o}gicas-Madrid, the University of Chicago, University College London, the DES-Brazil Consortium, the University of Edinburgh, 
the Eidgen{\"o}ssische Technische Hochschule (ETH) Z{\"u}rich, 
Fermi National Accelerator Laboratory, the University of Illinois at Urbana-Champaign, the Institut de Ci{\`e}ncies de l'Espai (IEEC/CSIC), 
the Institut de F{\'i}sica d'Altes Energies, Lawrence Berkeley National Laboratory, the Ludwig-Maximilians Universit{\"a}t M{\"u}nchen and the associated Excellence Cluster Universe, 
the University of Michigan, NFS's NOIRLab, the University of Nottingham, The Ohio State University, the University of Pennsylvania, the University of Portsmouth, 
SLAC National Accelerator Laboratory, Stanford University, the University of Sussex, Texas A\&M University, and the OzDES Membership Consortium.

Based in part on observations at Cerro Tololo Inter-American Observatory at NSF's NOIRLab (NOIRLab Prop. ID 2012B-0001; PI: J. Frieman), which is managed by the Association of Universities for Research in Astronomy (AURA) under a cooperative agreement with the National Science Foundation.

The DES data management system is supported by the National Science Foundation under Grant Numbers AST-1138766 and AST-1536171.
The DES participants from Spanish institutions are partially supported by MICINN under grants ESP2017-89838, PGC2018-094773, PGC2018-102021, SEV-2016-0588, SEV-2016-0597, and MDM-2015-0509, some of which include ERDF funds from the European Union. IFAE is partially funded by the CERCA program of the Generalitat de Catalunya.
Research leading to these results has received funding from the European Research
Council under the European Union's Seventh Framework Program (FP7/2007-2013) including ERC grant agreements 240672, 291329, and 306478.
We  acknowledge support from the Brazilian Instituto Nacional de Ci\^encia
e Tecnologia (INCT) do e-Universo (CNPq grant 465376/2014-2).

This manuscript has been authored by Fermi Research Alliance, LLC under Contract No. DE-AC02-07CH11359 with the U.S. Department of Energy, Office of Science, Office of High Energy Physics.

%Data Availability 
\section*{Data Availability}
The DES and OzDES data underlying this paper are available from \citet{DESDR2} and \citet{Lidman2020}, respectively. The lightcurves of the main sample sources are available in the online journal.

%%%%%%%%%%%%%%%%%%%%%%%%%%%%%%%%%%%%%%%%%%%%%%%%%%



\bibliographystyle{mnras}
\bibliography{ref} % if your bibtex file is called example.bib

%\begin{thebibliography}{99}
%\end{thebibliography}

\vspace{0.4cm}
\noindent \textbf{Affiliations}\\
$^{1}$Department of Astronomy, The Ohio State University, Columbus, Ohio 43210, USA\\
$^{2}$Center of Cosmology and Astro-Particle Physics, The Ohio State University, Columbus, Ohio, 43210, USA\\
$^{3}$Radcliffe Institute for Advanced Study, Harvard University, Cambridge, MA 02138, USA\\
$^{4}$School of Mathematics and Physics, University of Queensland, Brisbane, QLD 4072, Australia\\
$^{5}$ARC Centre of Excellence for All-sky Astrophysics (CAASTRO), 44 Rosehill Street Redfern, NSW 2016, Australia\\
$^{6}$Research School of Astronomy and Astrophysics, Australian National University, Canberra, ACT 2611, Australia\\
$^{7}$Australian Astronomical Observatory, North Ryde, NSW 2113, Australia\\
$^{8}$Space Telescope Science Institute, 3700 San Martin Drive, Baltimore, MD 21218, USA\\
$^{9}$Departamento de F\'isica Matem\'atica, Instituto de F\'isica, Universidade de S\~ao Paulo, CP 66318, S\~ao Paulo, SP, 05314-970, Brazil\\
$^{10}$Laborat\'orio Interinstitucional de e-Astronomia - LIneA, Rua Gal. Jos\'e Cristino 77, Rio de Janeiro, RJ - 20921-400, Brazil\\
$^{11}$Fermi National Accelerator Laboratory, P. O. Box 500, Batavia, IL 60510, USA\\
$^{12}$Instituto de F\'{i}sica Te\'orica, Universidade Estadual Paulista, S\~ao Paulo, Brazil\\
$^{13}$Centro de Investigaciones Energ\'eticas, Medioambientales y Tecnol\'ogicas (CIEMAT), Madrid, Spain\\
$^{14}$CNRS, UMR 7095, Institut d'Astrophysique de Paris, F-75014, Paris, France\\
$^{15}$Sorbonne Universit\'es, UPMC Univ Paris 06, UMR 7095, Institut d'Astrophysique de Paris, F-75014, Paris, France\\
$^{16}$Department of Physics \& Astronomy, University College London, Gower Street, London, WC1E 6BT, UK\\
$^{17}$SLAC National Accelerator Laboratory, Menlo Park, CA 94025, USA\\
$^{18}$Kavli Institute for Particle Astrophysics \& Cosmology, P. O. Box 2450, Stanford University, Stanford, CA 94305, USA\\
$^{19}$Instituto de Astrofisica de Canarias, E-38205 La Laguna, Tenerife, Spain\\
$^{20}$Universidad de La Laguna, Dpto. Astrofísica, E-38206 La Laguna, Tenerife, Spain\\
$^{21}$INAF, Astrophysical Observatory of Turin, Torino, Italy\\
$^{22}$Center for Astrophysical Surveys, National Center for Supercomputing Applications, 1205 West Clark St., Urbana, IL 61801, USA\\
$^{23}$Department of Astronomy, University of Illinois at Urbana-Champaign, 1002 W. Green Street, Urbana, IL 61801, USA\\
$^{24}$Astronomy Unit, Department of Physics, University of Trieste, via Tiepolo 11, I-34131 Trieste, Italy\\
$^{25}$INAF-Osservatorio Astronomico di Trieste, via G. B. Tiepolo 11, I-34143 Trieste, Italy\\
$^{26}$Institute for Fundamental Physics of the Universe, Via Beirut 2, 34014 Trieste, Italy\\
$^{27}$Observat\'orio Nacional, Rua Gal. Jos\'e Cristino 77, Rio de Janeiro, RJ - 20921-400, Brazil\\
$^{28}$Department of Physics, University of Michigan, Ann Arbor, MI 48109, USA\\
$^{29}$Santa Cruz Institute for Particle Physics, Santa Cruz, CA 95064, USA\\
$^{30}$Institute of Theoretical Astrophysics, University of Oslo. P.O. Box 1029 Blindern, NO-0315 Oslo, Norway\\
$^{31}$Kavli Institute for Cosmological Physics, University of Chicago, Chicago, IL 60637, USA\\
$^{32}$Instituto de Fisica Teorica UAM/CSIC, Universidad Autonoma de Madrid, 28049 Madrid, Spain\\
$^{33}$Institute of Space Sciences (ICE, CSIC),  Campus UAB, Carrer de Can Magrans, s/n,  08193 Barcelona, Spain\\
$^{34}$Institut d'Estudis Espacials de Catalunya (IEEC), 08034 Barcelona, Spain\\
$^{35}$Department of Astronomy, University of Michigan, Ann Arbor, MI 48109, USA\\
$^{36}$Department of Physics, Stanford University, 382 Via Pueblo Mall, Stanford, CA 94305, USA\\
$^{37}$Center for Astrophysics $\vert$ Harvard \& Smithsonian, 60 Garden Street, Cambridge, MA 02138, USA\\
$^{38}$Lawrence Berkeley National Laboratory, 1 Cyclotron Road, Berkeley, CA 94720, USA\\
$^{39}$Lowell Observatory, 1400 Mars Hill Rd, Flagstaff, AZ 86001, USA\\
$^{40}$Sydney Institute for Astronomy, School of Physics, A28, The University of Sydney, NSW 2006, Australia\\
$^{41}$Department of Physics and Astronomy, University of Pennsylvania, Philadelphia, PA 19104, USA\\
$^{42}$George P. and Cynthia Woods Mitchell Institute for Fundamental Physics and Astronomy, and Department of Physics and Astronomy, Texas A\&M University, College Station, TX 77843,  USA\\
$^{43}$Instituci\'o Catalana de Recerca i Estudis Avan\c{c}ats, E-08010 Barcelona, Spain\\
$^{44}$Institut de F\'{\i}sica d'Altes Energies (IFAE), The Barcelona Institute of Science and Technology, Campus UAB, 08193 Bellaterra (Barcelona) Spain\\
$^{45}$Physics Department, 2320 Chamberlin Hall, University of Wisconsin-Madison, 1150 University Avenue Madison, WI  53706-1390\\
$^{46}$Universite Clermont Auvergne, CNRS/IN2P3, LPC, F-63000 Clermont-Ferrand, France\\
$^{47}$Institute of Astronomy, University of Cambridge, Madingley Road, Cambridge CB3 0HA, UK\\
$^{48}$Department of Astrophysical Sciences, Princeton University, Peyton Hall, Princeton, NJ 08544, USA\\
$^{49}$School of Physics and Astronomy, University of Southampton,  Southampton, SO17 1BJ, UK\\
$^{50}$Computer Science and Mathematics Division, Oak Ridge National Laboratory, Oak Ridge, TN 37831\\
$^{51}$Institute of Cosmology and Gravitation, University of Portsmouth, Portsmouth, PO1 3FX, UK\\

%%%%%%%%%%%%%%%%%%%%%%%%%%%%%%%%%%%%%%%%%%%%%%%%%%

%%%%%%%%%%%%%%%%% APPENDICES %%%%%%%%%%%%%%%%%%%%%

\appendix

\section{Corner Plots for Continuum $+$ Iron Subtraction}

\begin{figure*}
\includegraphics[width=0.95\linewidth]{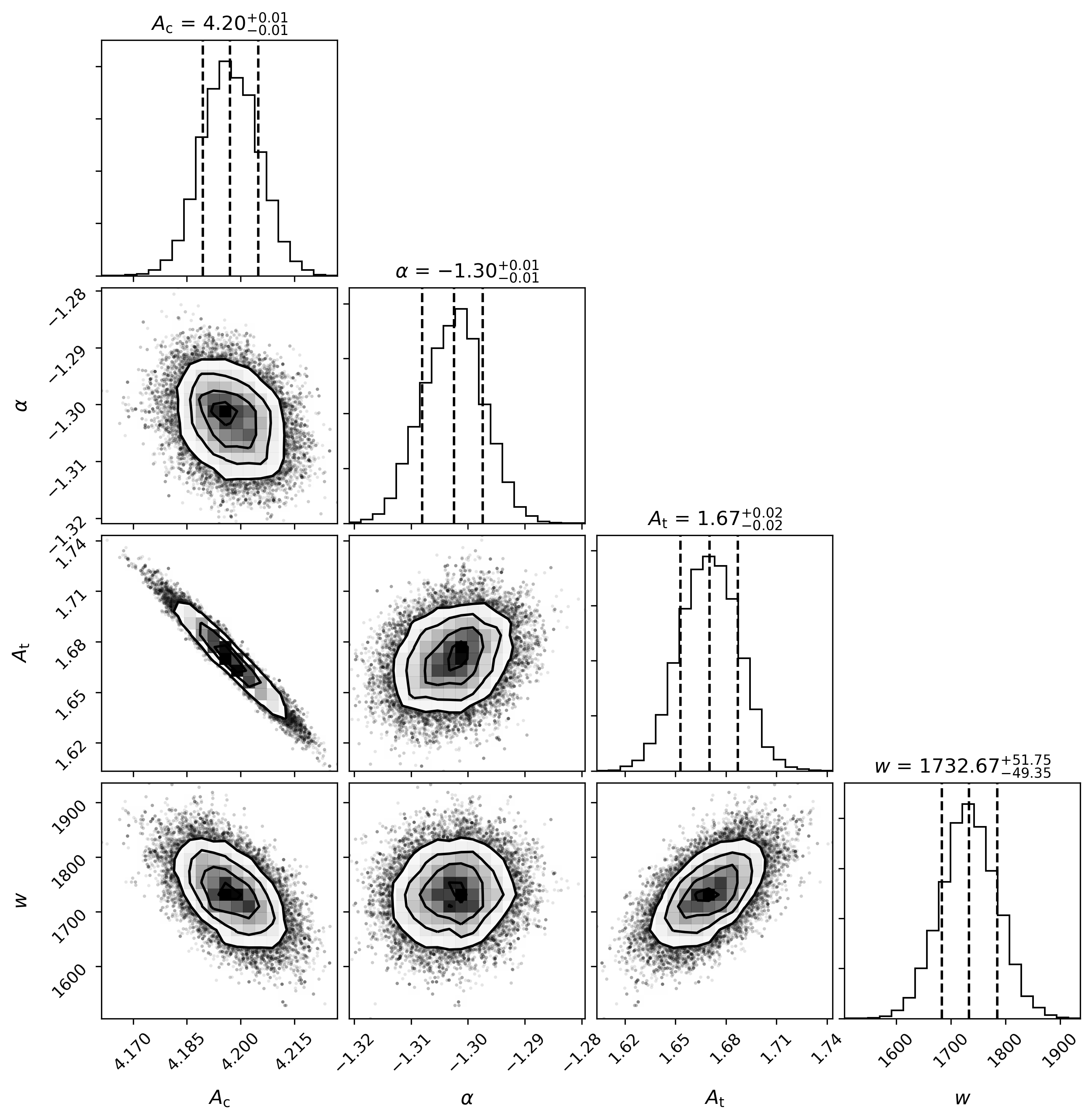}
\caption{Corner plot for the continuum $+$ iron fitting of the co-added spectra of the quasar DES J003052.76$-$430301.08 (middle panel of Figure \ref{fig:ironfit}). The iron template is from V01. The diagonal panels show the marginalized distributions of the four free parameters $A_{\rm c}$, $\alpha$, $A_{\rm t}$ and $w$, where $A_{\rm c}$ and $\alpha$ are the scale and slope of the power-law continuum, $A_{\rm t}$ is the scale of the iron template and $w$ is the velocity broadening width. The dashed lines mark the median and the 1$\sigma$ levels. The other panels show the correlations between parameters. The scatter points and the 2D histograms represent the MCMC chain steps and their distributions. The contours are drawn from the 0.5$\sigma$ to the $2\sigma$ level with a step of $0.5\sigma$.}
\label{fig:Corner_E0}
\end{figure*}

\begin{figure*}
\includegraphics[width=0.95\linewidth]{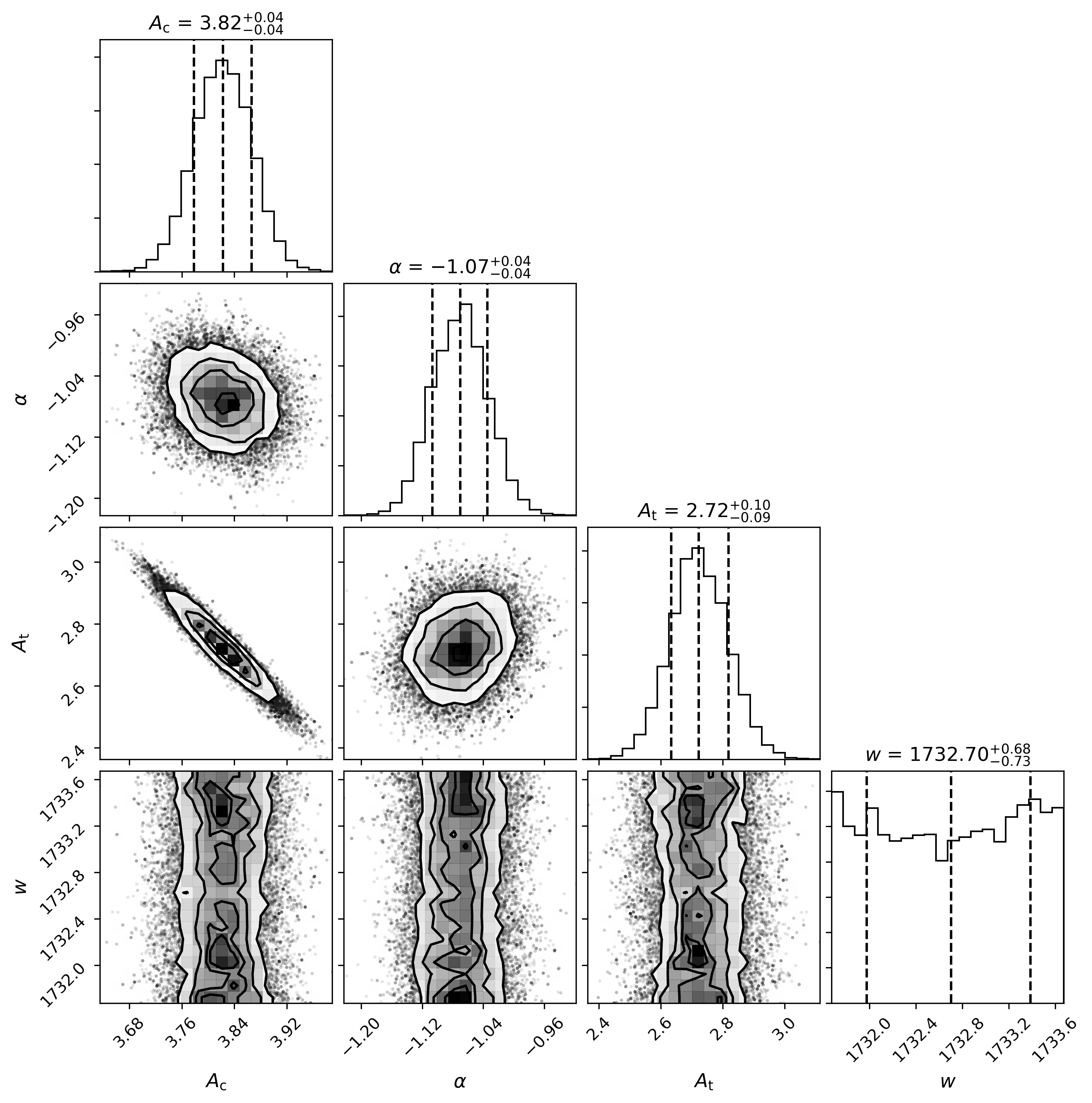}
\caption{Same as Figure \ref{fig:Corner_E0} but for the single-epoch spectra (bottom panel of Figure \ref{fig:ironfit}).}
\label{fig:Corner_E1}
\end{figure*}

Figures \ref{fig:Corner_E0} and \ref{fig:Corner_E1} show corner plots for the continuum $+$ iron fitting of the co-added spectra and the single-epoch spectra in Figure \ref{fig:ironfit}, respectively. The parameters show well-converged distributions. For the single-epoch spectra, fixing the velocity broadening width to an exact value would cause numerical problems in MCMC sampler \textsf{emcee} \citep{emcee}. We therefore allow $w$ to vary in a tiny range centered on the value derived from the co-added spectra. This range is small enough that it makes little difference that we did not fix the parameter. 

%%%%%%%%%%%%%%%%%%%%%%%%%%%%%%%%%%%%%%%%%%%%%%%%%%

% Don't change these lines
\bsp	% typesetting comment
\label{lastpage}
\end{document}